\documentclass[onecolumn,groupedaddress,showkeys]{article}
\usepackage{float}
\usepackage{graphicx}
\usepackage{amssymb}
\usepackage{authblk}
\usepackage{amsmath}
\usepackage{multirow}
\usepackage{hhline}
\usepackage{caption}
\usepackage{subcaption}
\usepackage{xcolor}
\usepackage{verbatim}
\usepackage[breaklinks]{hyperref}
\hypersetup{colorlinks,urlcolor=black,citecolor=black,linkcolor=black,filecolor=black}
\usepackage{breakurl}
\usepackage{anysize}
\usepackage[left=1cm,right=1cm,top=1cm,bottom=1cm,includehead,includefoot]{geometry}

\begin{document}

	\title{Simulating the spread of COVID-19 with cellular automata: A new approach}
	\author{Sourav Chowdhury\thanks{email: chowdhury95sourav@gmail.com}\qquad Suparna Roychowdhury\thanks{email: suparna@sxccal.edu}\qquad Indranath Chaudhuri\thanks{email: indranath@sxccal.edu}}
	\affil{Department of Physics, St. Xavier's College (Autonomous)\\
		30 Mother Teresa Sarani, Kolkata-700016, West Bengal, India}

	\maketitle
	
	\begin{abstract}	
		 Between the years 2020 to 2022, the world was hit by the pandemic of COVID-19 giving rise to an extremely grave situation. Many people suffered and died from this disease. Also the global economy was badly hurt due to the consequences of various intervention strategies (like social distancing, lockdown) which were applied by different countries to control this pandemic. There are multiple speculations that humanity will again face such pandemics in the future. Thus it is very important to learn and gain knowledge about the spread of such infectious diseases and the various factors which are responsible for it. In this study, we have extended our previous work (\textit{Chowdhury et.al., 2022}) on the probabilistic cellular automata (CA) model to reproduce the spread of COVID-19 in several countries by modifying its earlier used neighbourhood criteria. This modification gives us the liberty to adopt the effect of different restrictions like lockdown and social distancing in our model. We have done some theoretical analysis for initial infection and simulations to gain insights into our model. We have also studied the data from eight countries for COVID-19 in a window of 876 days and compared it with our model. We have developed a proper framework to fit our model on the data for confirmed cases of COVID-19 and have also  re-checked the goodness of the fit with the data of the deceased cases for this pandemic. Our model is compared with other well known CA models and the ODE based SEIR model. This model fits well with different peaks of COVID-19 data for all the eight countries and can be possibly generalized for a global prediction. Our study shows that the rate of disease spread depends both on infectivity of a disease and social restrictions. Also, it shows an overall decrement in mortality rate with time due to COVID-19 as more and more people get infected as well as vaccinated. Our minimal model with modified neighbourhood condition can easily quantify the degree of social restrictions. It is statistically concluded that the overall degree of social restrictions is above the mean when we considered all eight countries. Finally to conclude, this study has given us various important insights about this pandemic which can help in preparing for combating epidemics in future situations. 
	\end{abstract}
\section{Introduction}
	Epidemics and pandemics have been hampering human civilizations from centuries. In recent times, the COVID-19 disease had infected people of almost all countries globally. There are some claims that world will face a pandemic again in the near future \cite{future_1, future_2}. Thus it is very important to study and understand their nature. Mathematical models help us to simulate their behaviour and can give us many hidden details. 
	
	COVID-19 is caused by SARS-COV-2 virus. Globally, the number of infected cases and deaths from this disease is extremely high \cite{worldometers}. In India, this pandemic left a very serious toll on human life and economy. Due to rapid mutations, many variants of SARS-COV-2 have been found worldwide. Delta and Omicron are some commonly known mutations \cite{WHO}. Recently, many reports have been published about the worrying situation of COVID-19 in China \cite{NP_1, NP_2}.
	
	Epidemics and pandemics have been modeled through different mathematical and computational techniques. In the year 1927, Kermack and McKendrick established a dynamical model for epidemics known as the SIR model, which is a backbone for many current models \cite{basic_SIR}. This model consists of nonlinear differential equations. Many variants of this model like SEIR, SIRS and SEIAR have been developed throughout the year \cite{SIR_dengue, SIRS_influenza, Rep_No_Influenza, SEIAR_influenza, SEIR_ebola, Ebola_afr, cont_bio_model_measles}. These nonlinear differential equations based models have been extensively used to study COVID-19 pandemic \cite{covid_vac_india, covid_vac_acc_india, NPI_vs_vac, vac_resistant, HIT, 3rdwave_covid, chaos_vac, chaos_vac_osc}. Cellular automata is a spatio-temporal model which is represented with a collection of colored cells. The shape of these cells can be of any type like square and triangular \cite{chikungunya_vector_reg2}. Each of the color of the cells represent a state of the system. All the cells of the CA model evolve simultaneously and each cell evolves according to a predefined algorithm. This algorithm not only governs the interaction of a cell with neighbourhood cells but also controls evolution of the whole system. The states of the model and the neighbourhood condition depends on the system. In modeling epidemics, the four main states which have been considered are susceptible, exposed, infectious and recovered (or removed). Also, the Moore's and the Neumann's neighbourhood conditions are examples of trivial interaction methods which are used for CA to model epidemics. \cite{dengue_CA,Moore,Moore_2,Moore_3}
	
	CA has not only been used for diseases which spread by contact like influenza but also has been used in many vector-borne diseases like chikungunya, dengue \cite{ chikungunya_vector_reg2, influenza_seir_reg1, influenza_egypt, CA_influenza_abu_dhabi, influenzaA_SLEIRD, dengue_vector, dengue_CA, Ebola_CA}. Mostly in literature a disease spread is modeled in a square lattice where each lattice cell represents a person or a fraction of the population. However, there are some studies where a disease spread is modeled in a triangular or in a hexagonal lattice. In these models, the total population is divided into many sub-populations like Susceptible, Exposed, Infectious and Removed (SEIR) or various successors of it. In CA models, a neighbourhood condition is used to represent spatial interactions between cells. To study epidemics and pandemics, Moore’s neighbourhood condition and Neumann’s neighbourhood condition are the most widely used spatial interaction criteria. However these neighbourhood conditions lead to a serious issue which is called neighbourhood saturation \cite{CA_nghbd_sat}. This issue can be eliminated by taking a modified neighbourhood condition for example, Mikler et. al. proposed a global neighbourhood criteria to eliminate this issue \cite{CA_nghbd_sat, CA_small_world}. 
	
	Recently, cellular automata (CA) models have been used by many authors to study COVID-19 pandemic \cite{covid_social_isol, covid_vac_lockdwn, covid_GA_1, covid_mob_restr_net, COVID_GA_SC, CA_new_1, CA_new_2, CA_new_3}. In these studies CA has been mostly used to predict the future behaviour of the spread COVID-19 by studying its past behaviour. There are studies which also suggest various intervention and vaccination strategies to control this pandemic. Some advanced studies which have used the Genetic Algorithm (GA) and proposed various methods to optimize the CA model to fit the COVID-19 data \cite{covid_GA_1, covid_GA_2, COVID_GA_SC}. In these studies, different types of neighbourhood condition have been used. A particular neighbourhood condition called the $r$-neighbourhood condition is used widely in literature \cite{covid_vac_lockdwn, covid_GA_1}. In this model a person can interact upto the $r$th neighbours with equal probabilities and thus increasing the range of disease spread beyond nearest neighbours.
	
	In this study, we have defined a generalized neighbourhood condition. We have analyzed and studied this neighbourhood condition in detail to get a clear picture of its behaviour. We have applied this condition in our CA model and have done some analytical estimates. Various simulations have been done on this system after defining a suitable algorithm. We have also compared our model with other current models used in literature. Finally, we have studied COVID-19 data from eight different countries in relevance to our model and analyzed the concurrence between the two. 
	
	This paper is organized as follows: In Sec.~\ref{sec2}, we have defined our neighbourhood condition. An initial analysis on this neighbourhood condition has been done in Sec.~\ref{sec3}. In Sec.~\ref{prb_inf} and Sec.~\ref{an_study}, probability of infection has been calculated and theoretical analysis of our CA model is described respectively. In Sec.~\ref{algo_sim}, we have defined an algorithm for this CA model and shown different results of our simulations. Our model is compared to other models in Sec.~\ref{cont_comp_sec} and \ref{ngh_comp_sec}. The detailed study on COVID-19 data using our CA model is described in Sec.~\ref{cov_fit_sec}. Finally we have listed the concluding remarks in Sec.~\ref{conclusion}.

\section{Model description}\label{sec2}
	In this article, we have studied the SEIR model by cellular automata with a modified neighbourhood condition. In this section, we mainly discuss our neighbourhood condition in detail and also its features which will affect a real situation. All the other important assumptions for this model are similar to Chowdhury \textit{et. al} (2022) \cite{COVID_GA_SC} where the choice of periodic boundary condition is of special importance in this study. 
	\subsection{Neighbourhood condition}
\begin{figure}[H]
		\centering
		\includegraphics[scale=0.55]{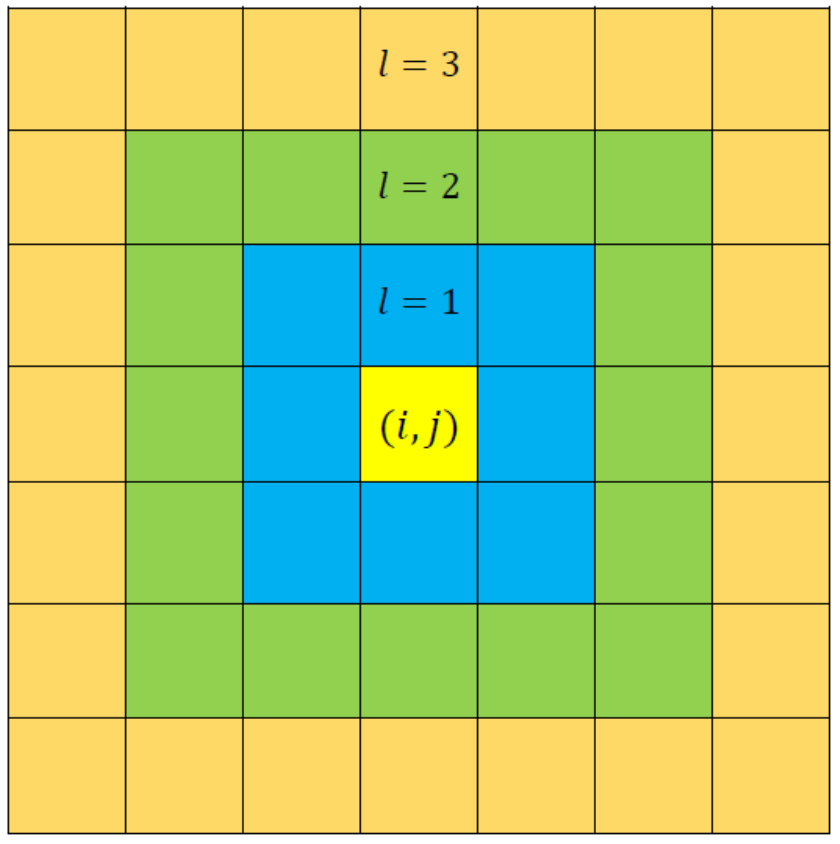}
		\caption{{Different layers of neighbourhoods corresponding to the cell, $(i,j)$ of a lattice. \cite{COVID_GA_SC}.}\label{nghbd_custom}}
	\end{figure}

	Suppose the lattice size for this CA model is $N\times N$. Then with respect to any $(i,j)$ cell, the lattice can be divided into several layers (by considering periodic boundary condition) as shown in Fig.~\ref{nghbd_custom}. 
	
	Here $\ell$ represents the layer number. Any layer of number $\ell$ has $8\ell$ number of cells. With respect to any cell, indexed by $(i,j)$, of a $N\times N$ lattice, the total number of layers can be calculated by
	
	\begin{equation}
		L=\frac{N-1}{2} \label{L_rel}
	\end{equation} 
	
	when $N$ is odd. If, $N$ is even and $N\gg 1$ then Eq.\ref{L_rel} is approximately true. The main assumptions for this modified neighbourhood condition, similar to Chowdhury \textit{et. al} (2022) and references there in \cite{COVID_GA_SC} are
	
	\begin{itemize}
		\item The distance between any two cells is proportional to the layer number of any one cell with respect to the other cell. Hence, $d\propto \ell$. For, simplicity we can assume $d=\ell$.
		\item Probability of interaction ($p_{\scriptscriptstyle{\text{int}}}(d)$) between two cells is inversely proportional to the $n$th power of the distance. Thus, $p_{\scriptscriptstyle{\text{int}}}(d)\propto \dfrac{1}{d^{n}}$. Here $n$= degree exponent and it can have any positive or negative values.
	\end{itemize}
	
	Thus probability of interaction ($p_{\scriptscriptstyle{\text{int}}}(d)$) can be written as,
	
	\begin{equation}
		p_{\scriptscriptstyle{\text{int}}}(d)=\frac{ \frac{1}{d^{n}}}{\sum_{d}\frac{1}{d^{n}}}=\frac{ \frac{1}{\ell^{n}}}{\sum_{\ell=1}^{L}\frac{1}{\ell^{n}}}=\frac{1}{A_{n}\ell^{n}}\label{pint_rel}
	\end{equation} 

	where, $A_{n}=\sum_{\ell=1}^{L}\frac{1}{\ell^{n}}$. 
	
	\subsection{Average distance of interaction ($\langle d\rangle$)}\label{sec3}
	In modeling disease spread, the interaction between people is very important. In real world, interaction between two persons will depend on the distance between them. As distance increases, chance of interaction between two persons decrease \cite{sp_1,sp_4}. Also at the time of COVID-19 pandemic, interactions between people were controlled by various restrictions like lockdown and social distancing. These facts motivate us to assume the power law form of the interaction probability as shown in Eq.~\ref{pint_rel}. Here degree exponent $n$ can tune the interaction probability as per the requirements of different strictness of restrictions on social interactions and models.
	
	In this section, we have defined the average distance of interaction and discuss its dependence on the degree exponent $n$. The average distance of interaction ($\langle d\rangle$) and the variance of the interaction distance ($\sigma^{2}_{d}$) can be deduced from Eq.\ref{pint_rel} as,
	
	\begin{equation}
		\langle d\rangle=\sum_{\ell=1}^{L}\ell p_{\scriptscriptstyle{\text{int}}}(\ell)=\frac{1}{A_{n}}\sum_{\ell=1}^{L}\frac{1}{\ell^{n-1}}
	\end{equation}

	\begin{equation}
		\sigma_{d}^{2}=\langle d^{2}\rangle-\langle d\rangle^{2}= \frac{1}{A_{n}}\sum_{\ell=1}^{L}\frac{1}{\ell^{n-2}}-\left(\frac{1}{A_{n}}\sum_{\ell=1}^{L}\frac{1}{\ell^{n-1}}\right)^{2}
	\end{equation}
	
				\begin{figure}[H]
		\begin{subfigure}{.5\textwidth}
			\centering
			\includegraphics[scale=0.65]{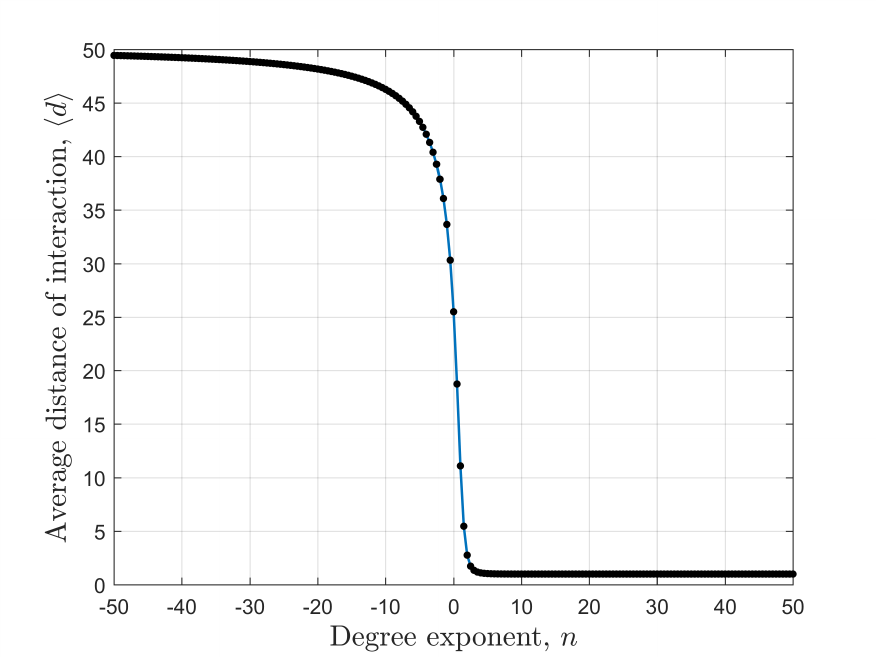}
			\caption{}
			\label{avg_int_n}
		\end{subfigure}
		\begin{subfigure}{.5\textwidth}
			\centering
			\includegraphics[scale=0.65]{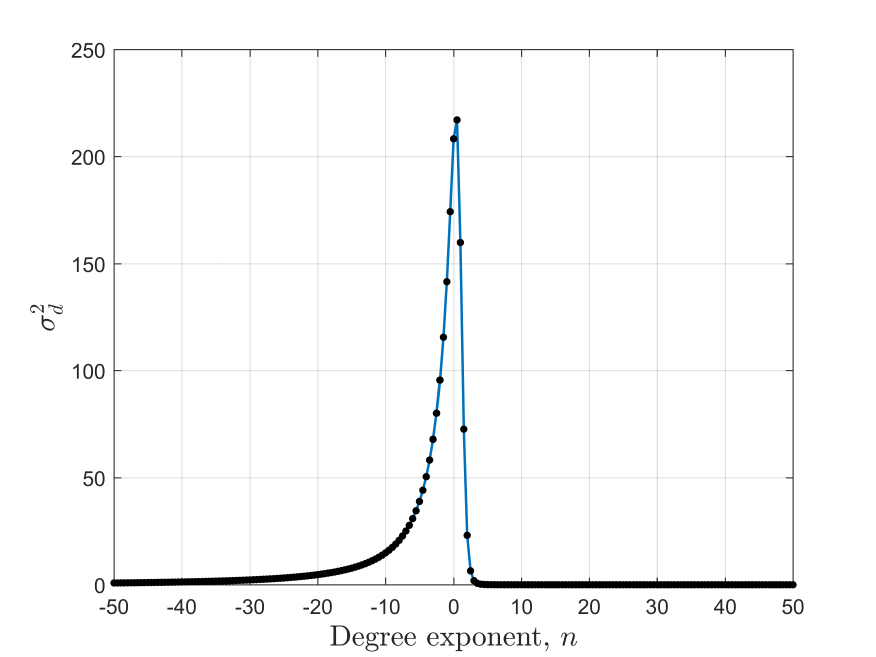}
			\caption{}
			\label{Var_n}
		\end{subfigure}
		\caption{{Average distance of interaction ($\langle d\rangle$) (a) and variance on the distance ($\sigma_{d}^2$) (b) are plotted as a function of $n$. Here we have considered $L=50$.}\label{avg_int}}	
	\end{figure}
	
	Fig.~\ref{avg_int_n} shows that the average distance of interaction ($\langle d\rangle$) falls very rapidly when degree exponent ($n$) changes from negative to positive. Also, Fig.~\ref{Var_n} shows that variance ($\sigma_{d}^{2}$) has a very sharp peak for $n$ which have slightly higher value than zero. Numerically it is found that $\sigma_{d}^{2}$ has a peak at $n\approx 0.3519$ for $L=50$. So, the $n$ values around 0.0 are very significant because of the high values of $\sigma_{d}^{2}$. When $n$ is positive then $\langle d\rangle$ quickly saturates to one and for negative values of $n$, $\langle d\rangle$ quickly saturates to maximum layer number ($L$). Variance of the interaction distance, $\sigma_{d}^{2}$ rapidly saturates to zero for both positive and negative values of $n$. As, $\langle d\rangle$ decreases with increase in $n$, $n$ can be defined as `social confinement' \cite{COVID_GA_SC}. Also, we want to mention that a high negative value incorporates that a person will mainly interact with the distant persons which is physically not possible. Thus very high positive or negative values of the degree exponent is physically irrelevant.

	\section{Probability of infection}\label{prb_inf}
	Probability of infection of any susceptible cell at $(i,j)$ ($Q_{\text{I}}(i,j)$) states the total probability of infection of this susceptible cell due to the infectious cells around it. To calculate $Q_{\text{I}}(i,j)$ it is needed to define some probabilities which are given below:
	
	\begin{itemize}
		\item Probability of infection per contact ($q$): This defines the probability of infection when a susceptible cell interacts with an infectious cell. This probability depends on the infectivity of the virus of a disease and the immunity of a person. If a virus is more infectious than its previous mutations, the probability of infection per contact will increase. Also, if a population gains immunity from a disease, then the probability of infection per contact will decrease. For COVID-19, the immunity is mainly achieved by vaccination.    
		
		\item Probability that an infectious person is at a distance $d$ is given by ($p_{\scriptscriptstyle{\text{I}}}(d)$): $p_{\scriptscriptstyle{\text{I}}}(d)$ can be represented as,
		
		\begin{equation}
			p_{\scriptscriptstyle{\text{I}}}(d)=\frac{\text{Number of infectious cells in layer}\ \ell}{\text{Total number of cells in layer}\ \ell}=\frac{N_{\text{I}}(\ell)}{8\ell} \label{pI_rel}
		\end{equation}
	\end{itemize}

	 Hence from these probabilities the probability of infection ($Q_{I}(i,j)$) of any susceptible cell at $(i,j)$ can be derived as,
	 
 	\begin{equation}
	 Q_{\text{I}}= \sum_{d} qp_{\scriptscriptstyle{\text{int}}}(d)p_{\scriptscriptstyle{\text{I}}}(d)=q \sum_{\ell=1}^{L} p_{\scriptscriptstyle{\text{int}}}(\ell)p_{\scriptscriptstyle{\text{I}}}(\ell). \label{prob_inf1}
	 \end{equation}	
 
	 Substituting Eq.~\ref{pint_rel} and Eq.~\ref{pI_rel} in Eq.~\ref{prob_inf1} we get,
	 
	 \begin{equation}
	 	Q_{\text{I}}=\frac{q}{8A_{n}}\sum_{\ell=1}^{L}\frac{N_{\text{I}}(\ell)}{\ell^{n+1}} \label{prob_inf2}
	 \end{equation}	
 	
	So, it is noted that the probability of infection ($Q_{\text{I}}(i,j)$) is highly depended on the degree exponent $n$. If $n\gg 1$ then probability of infection of any susceptible cell depends on the number nearest infectious cells only.
	
\section{Dependence of initial infection growth on various free parameters}\label{an_study}
	In this section, we have tried to find the initial infection growth of our model and its dependence on various free parameters. Here we have assumed that initially there are no removed persons in the region, which means $R\!\left(0\right)=0$.
	\subsection{Initial infection growth with one initial infection}
	Let, the total size of the lattice is $N\times N$. Also, assuming that at time, $t=0$ there is only one infectious person in the region. This infectious person has $8\ell$ number of susceptible neighbours at a distance $\ell$, which can be understood from Fig.~\ref{nghbd_custom}. In other words, there are $8\ell$ susceptible persons who are placed at a distance $\ell$ from the infectious person. Thus the probability of infection for each of these $8\ell$ susceptible persons is 
	\begin{equation}
		Q_{\text{I}}(\ell)=\frac{q}{8A_{n}}\frac{1}{\ell^{n+1}}
	\end{equation}
	Hence, on average, the number of newly infected cases between these $8\ell$ susceptible persons at $t=1$ is,
	\begin{equation}
		E_{\ell}(1)=8\ell\left(\frac{q}{8A_{n}}\frac{1}{\ell^{n+1}}\right)=\frac{q}{A_{n}}\frac{1}{\ell^{n}}
	\end{equation}
	Thus considering all distances, the average number of newly infected cases at $t=1$ are, 
	\begin{equation}
		E(1)=\sum_{\ell=1}^{L}E_{\ell}(1)=\frac{q}{A_{n}}\sum_{\ell=1}^{L}\frac{1}{\ell^{n}}=\frac{q}{A_{n}}A_{n}=q
	\end{equation}
	So, with one initial infectious case, the average number of newly infected cases at $t=1$ is a constant and have a value equal to $q$, which is the probability of infection per contact and is independent of $n$.
	\subsection{New infected cases with multiple initial infections}
	Let at $t=0$, there are $n_{\scriptscriptstyle{\text{I}}}$ number of infectious cells and which are represented by $i_{1},i_{2},...,i_{n_{\scriptscriptstyle{\text{I}}}}$ and the distance between any two infectious cells $i_{p}$ and $i_{q}$ is represented by $\ell_{pq}$. To estimate the new infections, we will consider each infectious person one by one. An infectious person can infect everyone except the already infected person. Thus, the average number of susceptible persons who can only be infected by $i_{1}$ is,
	\begin{equation}
		E_{i_{1}}(1)=q-\frac{q}{8A_{n}}\left(\frac{1}{\ell_{12}^{n+1}}+\frac{1}{\ell_{13}^{n+1}}+...+\frac{1}{\ell_{1n_{\scriptscriptstyle{\text{I}}}}^{n+1}}\right)\label{Ei1}
	\end{equation} 
	As an infectious person cannot infect another infectious person, we have subtracted those probabilities from $q$. Considering all infectious cases, we can write,
	\begin{table}[H]
		\centering
		\begin{tabular}{l}
			$E_{i_{1}}(1)=q-\frac{q}{8A_{n}}\left(\frac{1}{\ell_{12}^{n+1}}+\frac{1}{\ell_{13}^{n+1}}+...+\frac{1}{\ell_{1n_{\scriptscriptstyle{\text{I}}}}^{n+1}}\right)$\\
			$E_{i_{2}}(1)=q-\frac{q}{8A_{n}}\left(\frac{1}{\ell_{21}^{n+1}}+\frac{1}{\ell_{23}^{n+1}}+...+\frac{1}{\ell_{2n_{\scriptscriptstyle{\text{I}}}}^{n+1}}\right)$\\
			.\\
			.\\
			.\\
			$E_{i_{n_{\scriptscriptstyle{\text{I}}}}}(1)=q-\frac{q}{8A_{n}}\left(\frac{1}{\ell_{n_{\scriptscriptstyle{\text{I}}}1}^{n+1}}+\frac{1}{\ell_{n_{\scriptscriptstyle{\text{I}}}2}^{n+1}}+...+\frac{1}{\ell_{n_{\scriptscriptstyle{\text{I}}}n_{\scriptscriptstyle{\text{I}}}-1}^{n+1}}\right)$\\	
			\hline
			$E(1)=n_{\scriptscriptstyle{\text{I}}}q-\frac{2q}{8A_{n}}\left(\sum\limits_{\substack{i,j=1 \\ i<j}}^{n_{\scriptscriptstyle{\text{I}}}}\frac{1}{\ell_{ij}^{n+1}}\right)$
		\end{tabular}
	\end{table}
	As $\ell_{ij}=\ell_{ji}$, every term in the summation repeats twice. Hence the new infected cases at time $t=1$ are,
	\begin{equation}
		E(1)=n_{\scriptscriptstyle{\text{I}}}q-\frac{2q}{8A_{n}}\left(\sum_{\substack{i,j=1 \\ i<j}}^{n_{\scriptscriptstyle{\text{I}}}}\frac{1}{\ell_{ij}^{n+1}}\right) \label{new_multi}
	\end{equation}
	The summation of Eq.~\ref{new_multi} consists of $^{n_{\scriptscriptstyle{\text{I}}}}\!C_{2}$ number of terms. If $n_{\scriptscriptstyle{\text{I}}}\ll N^{2}$,  $^{n_{\scriptscriptstyle{\text{I}}}}\!C_{2}$ number of relative terms in Eq.~\ref{new_multi} are negligible. Thus we can write,
	\begin{equation}
		E(1)\approx n_{\scriptscriptstyle{\text{I}}}q \label{new_multi_approx}.
	\end{equation}
	 Next we will look at another method of estimating average number of newly infected cases with randomly distributed initial infections.
	 We note that the relative terms in Eq.~\ref{new_multi} are difficult to estimate as they can have different values for different simulation as a function of relative distances. However, we can adopt an average approach to this. Thus from Eq.~\ref{Ei1}, the average new infected cases for $i_{1}$ can be written as,
	\begin{equation}
		\overline{E_{i_{1}}(1)}=q-\frac{q}{8A_{n}}\overline{\sum_{j=2}^{n_{I}}\frac{1}{\ell_{1j}^{n+1}}}=q-\frac{q}{8A_{n}}\sum_{j=2}^{n_{I}}\overline{\frac{1}{\ell_{1j}^{n+1}}} \label{Ei1_avg_prlim}
	\end{equation} 
	Probability that an infectious person will be randomly placed in layer $\ell$ with respect to $i_{1}$ is $\frac{8\ell}{N^{2}-1}$. Here, we take $N^{2}-1$ instead of $N^{2}$ as one place is occupied by $i_{1}$. Hence,
	\begin{equation}
		\overline{\frac{1}{\ell_{1j}^{n+1}}}=\sum_{\ell_{1j}=1}^{L}\frac{1}{\ell_{1j}^{n+1}}\frac{8\ell_{1j}}{N^{2}-1}=\frac{8}{N^{2}-1}\sum_{\ell_{1j}=1}^{L}\frac{1}{\ell_{1j}^{n}}=\frac{8A_{n}}{N^{2}-1}\label{lij}
	\end{equation} 
	All $n_{\scriptscriptstyle{\text{I}}}-1$ terms in the sum given in Eq.~\ref{Ei1_avg_prlim} have the same average as given in Eq.~\ref{lij}. Hence, from Eq.~\ref{Ei1_avg_prlim} we can write,
	\begin{equation}
		\overline{E_{i_{1}}(1)}=q-\frac{q}{N^{2}-1}\left(n_{\scriptscriptstyle{\text{I}}}-1\right)
	\end{equation} 
	Hence, for all infectious cases,
	\begin{equation}
		\overline{E(1)}=n_{\scriptscriptstyle{\text{I}}}\overline{E_{i_{1}}(1)}=n_{\scriptscriptstyle{\text{I}}}q-\frac{qn_{\scriptscriptstyle{\text{I}}}\left(n_{\scriptscriptstyle{\text{I}}}-1\right)}{N^{2}-1}
	\end{equation} 
	\subsection{Interpreting the basic reproduction number}
	Let, $\tau_{\scriptscriptstyle{\text{R}}}$ represent the mean infectious period. If initial infectious cases, $n_{\scriptscriptstyle{\text{I}}}\ll N^{2}$, the new infected cases, $E(1)\approx n_{\scriptscriptstyle{\text{I}}}q$ (Eq.~\ref{new_multi_approx}) after unit time steps. So, one infectious person will spread approximately $q$ number of infections. As the infectious period is $\tau_{\scriptscriptstyle{\text{R}}}$ one infectious person will spread $q\tau_{\scriptscriptstyle{\text{R}}}$ number of infections in this period. This quantity is equivalent to the basic reproduction number ($R_{0}$).

\section{Algorithm and simulation of the CA model}\label{algo_sim}
	In this section, we have discussed about the algorithm of our CA model and shown multiple simulation results. 
	\subsection{Algorithm}\label{algo}
		This CA model consists of four sub-populations: susceptible ($S$), exposed ($E$), infectious ($I$) and removed ($R$) (recovered+dead). At any time $t$, each cell of the lattice belongs to any one of these sub-populations. For simulations, these sub-populations are represented with values 0, 1, 2 and 3 respectively. The algorithm for the CA model is same as described in Chowdhury \textit{et. al} (2022) \cite{COVID_GA_SC}. 
		
		Suppose, the lattice size is $N\times N$ and at time $t$ the number of cells belonging to different sub-populations like exposed, infectious and removed are $N_{\text{E}}=E\left(t\right)$, $N_{\text{I}}=I\left(t\right)$, and $N_{\text{R}}=R\left(t\right)$ respectively. Thus the number of cells belonging to the susceptible sub-population at time $t$ is $S(t)=N^{2}-N_{\text{E}}-N_{\text{I}}-N_{\text{R}}$. Any susceptible cell at position $(i,j)$ can be infected with a probability $Q_{\text{I}}\left(i,j,t\right)$ as described in Eq.~\ref{prob_inf2}. These newly infected cells at time $t$ will be moved to the exposed sub-population at $t+1$. Also, the exposed and infectious cells will move to the infectious and removed sub-populations respectively after the latency period ($\tau_{\scriptscriptstyle{\text{I}}}$) and infectious period ($\tau_{\scriptscriptstyle{\text{R}}}$) for the disease. Therefore at time $t$, if a cell is exposed (or infectious), it will move to the infectious (or removed) sub-population at time $t+\tau_{\scriptscriptstyle{\text{I}}}$ (or $t+\tau_{\scriptscriptstyle{\text{R}}}$).   
		
	\subsection{Simulated results}\label{sim_sec}
		In this section, we have discussed various simulations and results of our model. To start with, all the free parameters of our simulation are lattice size ($N\times N$), sample size ($S$), probability of infection per contact ($q$), degree exponent ($n$), mean latency period of the disease ($\tau_{\scriptscriptstyle{\text{I}}}$), and mean infectious period of the disease ($\tau_{\scriptscriptstyle{\text{R}}}$).
	
		For our simulations, we have assumed the values of  $\tau_{\scriptscriptstyle{\text{I}}}$ and $\tau_{\scriptscriptstyle{\text{R}}}$ as  $\tau_{\scriptscriptstyle{\text{I}}}=8~\text{days}$ and $\tau_{\scriptscriptstyle{\text{R}}}=18~\text{days}$. Also we have assumed that the initially there is only one infectious cell in the whole lattice and all other cells are susceptible. Hence, the initial condition for the CA model with a $N\times N$ lattice is $I\left(0\right)=1$, $E\left(0\right)=0$, $R\left(0\right)=0$, $S\left(0\right)=N^{2}-1$.
	
\begin{figure}[H]
		 	\begin{subfigure}{.5\textwidth}
		 		\centering
		 		\includegraphics[scale=0.65]{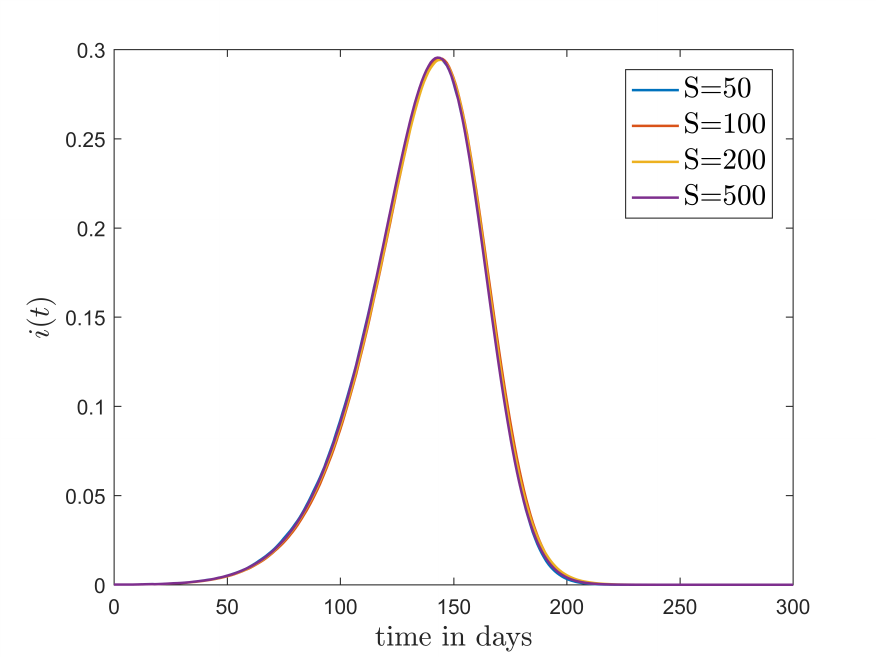}
		 		\caption{}
		 		\label{sample_var}
		 	\end{subfigure}
		 	\begin{subfigure}{.5\textwidth}
		 		\centering
		 		\includegraphics[scale=0.65]{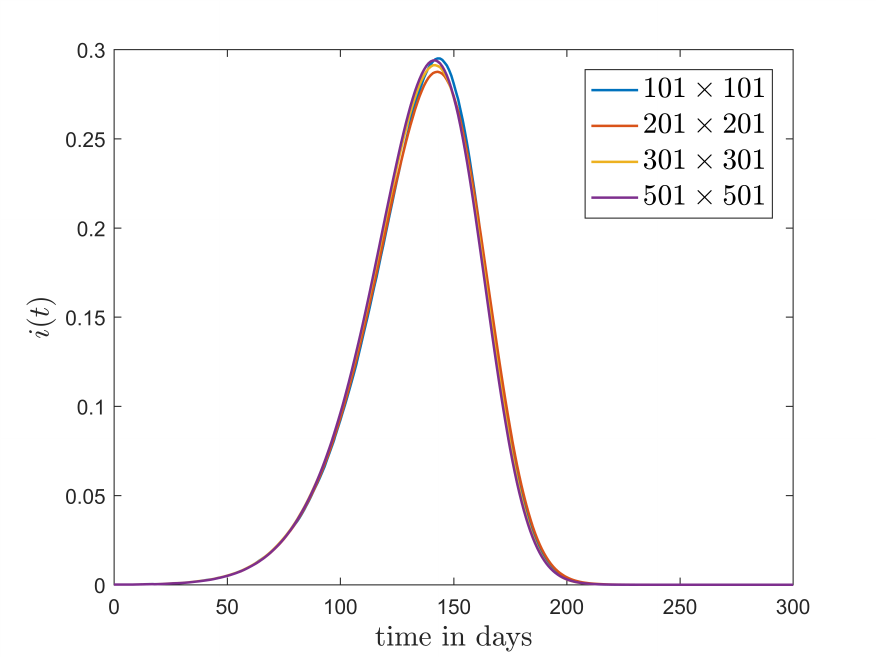}
		 		\caption{}
		 		\label{size_var}
		 	\end{subfigure}
		 	\caption{Plots of the variations of $i(t)$ for different sample and lattice sizes. (a) Plot of the variations of $i(t)$ for different sample sizes when lattice size fixed at $101\times 101$. (b) Plot of the variations of $i(t)$ for different lattice sizes when sample size fixed at 50.}\label{sample_and_size}
		 \end{figure}

		 At first, we have checked the dependence of the model on the two free parameters, sample size ($S$) and lattice size ($N\times N$). We have checked this through the time evolution of the fraction of the infectious cases ($i(t)=\frac{I(t)}{N^{2}}$). The results are shown in the Fig.~\ref{sample_and_size}. 
		 
	 	Fig.~\ref{sample_var} shows the variation of $i(t)$ which is averaged out for different sample sizes when $q=0.3$, $n=2$ and lattice size is fixed at $101\times 101$.  It is found that there is no major deviation happens in $i(t)$ by changing sample size.
	 	
	 	Fig.~\ref{size_var} shows the variation of $i(t)$ for different lattice sizes. Here, we have again considered $q=0.3$, $n=2$ and also the sample size as $S=50$.  It is found that if the initial fraction of the infection cases ($i(0)$) is same then the evolution of $i(t)$ is approximately independent of the lattice size. Here we have checked the results with different combinations of the parameters for both of these cases. Thus we can conclude, the model is approximately sample and size independent. For future simulations we have fixed the lattice size at $N\times N=101\times 101$ and sample size at $S=50$.
	 	
	 	\begin{figure}[H]
	 		\begin{subfigure}{.5\textwidth}
	 			\centering
	 			\includegraphics[scale=0.65]{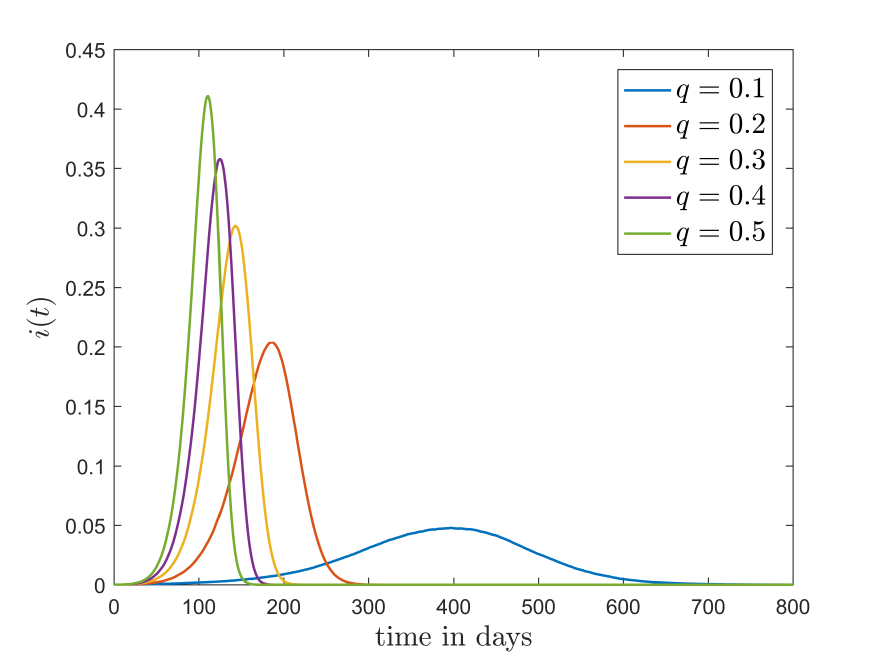}
	 			\caption{}
	 			\label{q_var}
	 		\end{subfigure}
	 		\begin{subfigure}{.5\textwidth}
	 			\centering
	 			\includegraphics[scale=0.65]{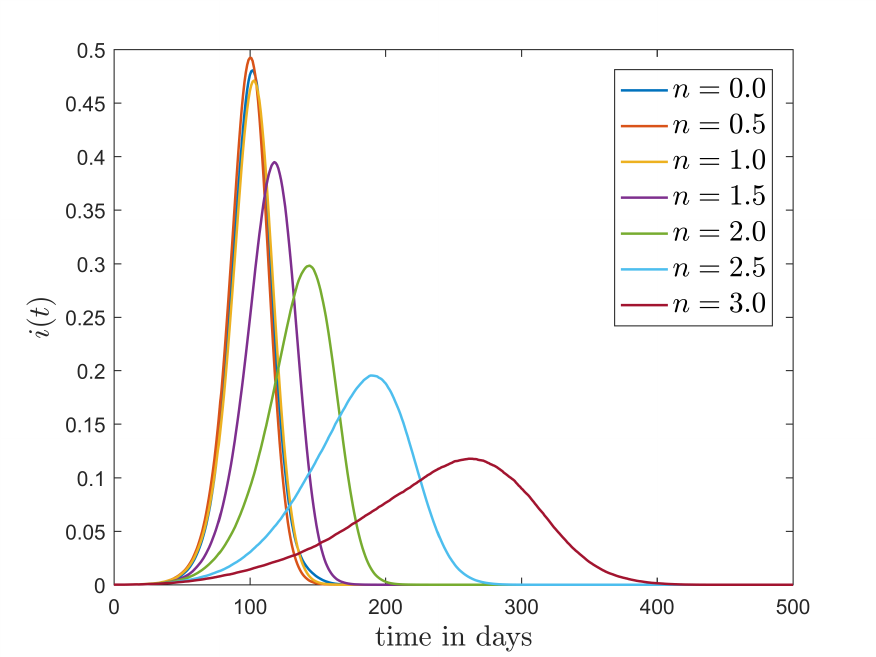}
	 			\caption{}
	 			\label{n_var}
	 		\end{subfigure}
	 		\caption{Plots of $i(t)$ for different values of $q$ and $n$ (a) Plot of $i(t)$ for different values of $q$ for $n=2.0$. (b) Plot of $i(t)$ for different values of $n$ for $q=0.3$. }\label{q_and_n_var}	
	 	\end{figure}
	 	Fig.~\ref{q_and_n_var} shows how the the fraction of the infectious cases ($i(t)$) varies with time for different $q$ values and $n$ values. In Fig.\ref{q_var} it is shown how $i(t)$ varies for different $q$ when $n=2$. Here, we can see that growth of $i(t)$ is very fast for high $q$ values. This is reasonable because high $q$ means when a susceptible people cell with an infectious cell then there is a very high chance that the susceptible people will be infected. Thus the disease spread is faster for high $q$ values than the low $q$ values. 
	 	
	 	To simulate Fig.~\ref{n_var} we have fixed $q$ at $q=0.3$. It can be seen from this figure that when $n$ increases the peak of the fraction of the infectious cases ($i(t)$) falls significantly and disease spread become slower.
	 	
	 	\begin{figure}[H]
	 		\centering
	 		\begin{tabular}{c c c c}
	 			\includegraphics[scale=0.3]{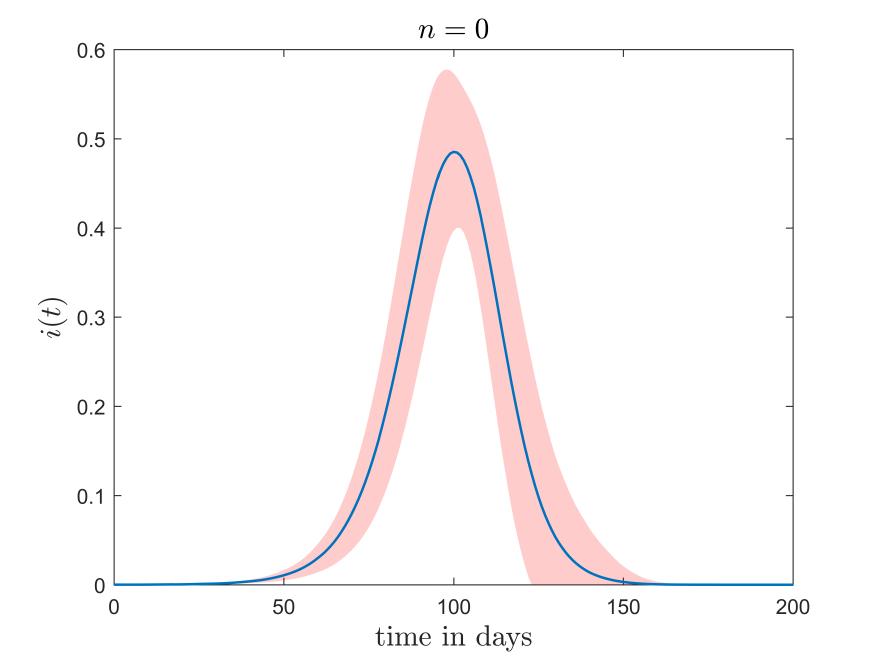}& 
	 			\includegraphics[scale=0.3]{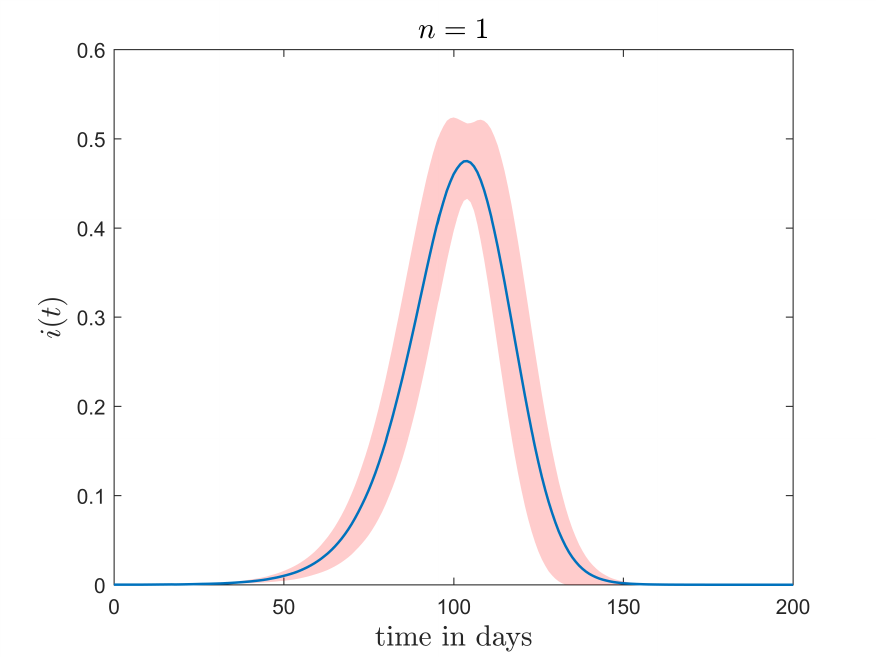}& 
	 			\includegraphics[scale=0.3]{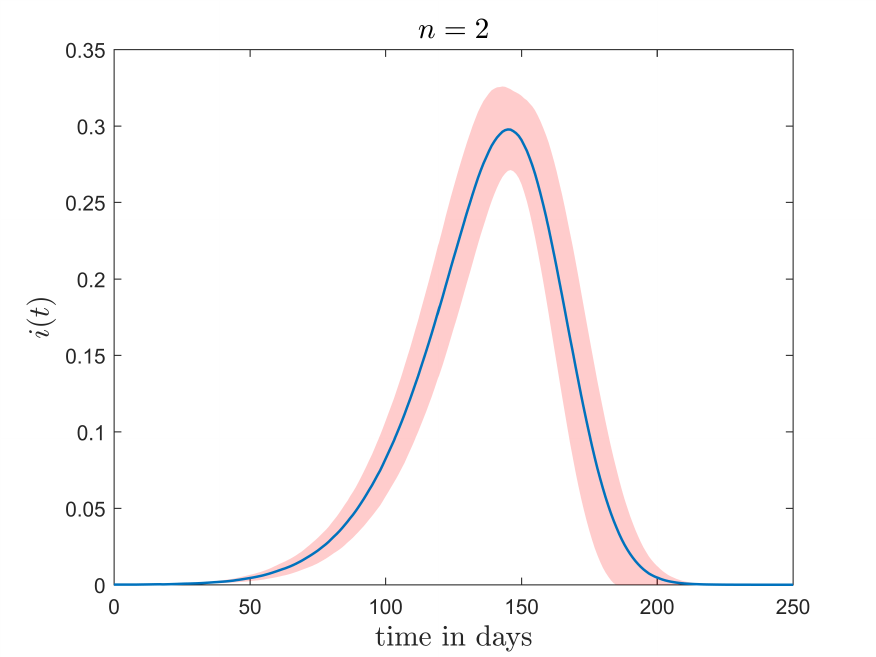}&
	 			\includegraphics[scale=0.3]{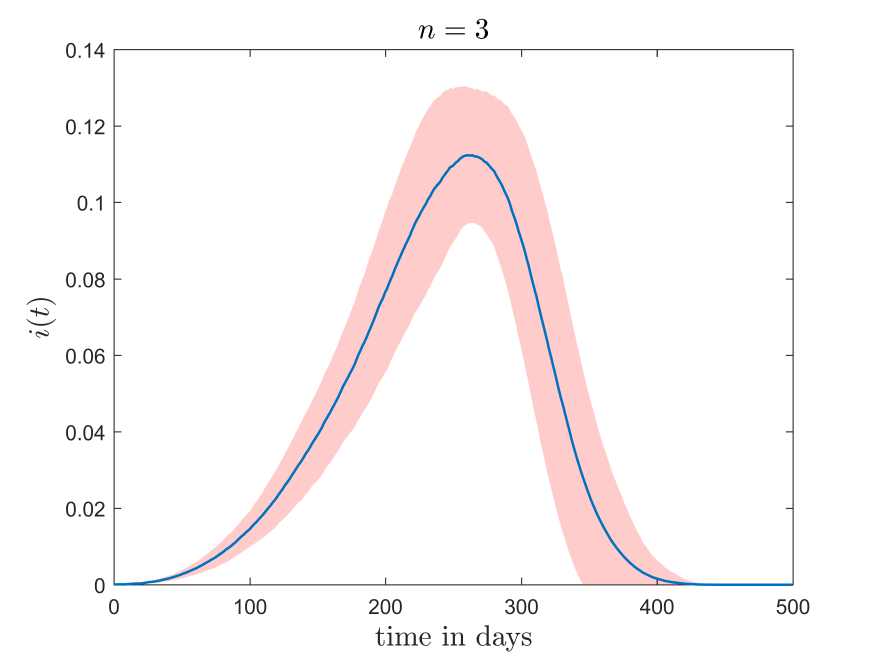}\\
	 			\includegraphics[scale=0.3]{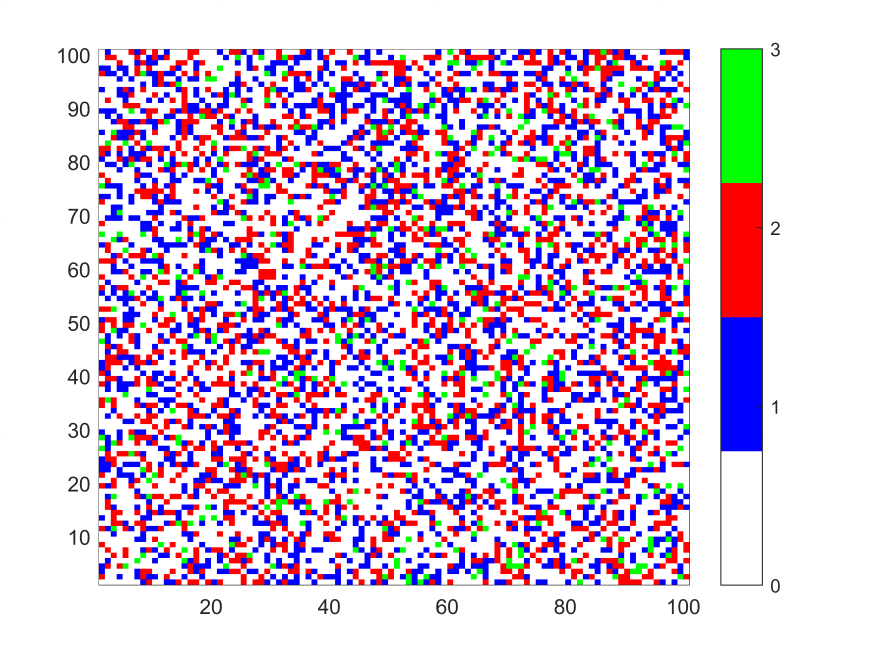}& 
	 			\includegraphics[scale=0.3]{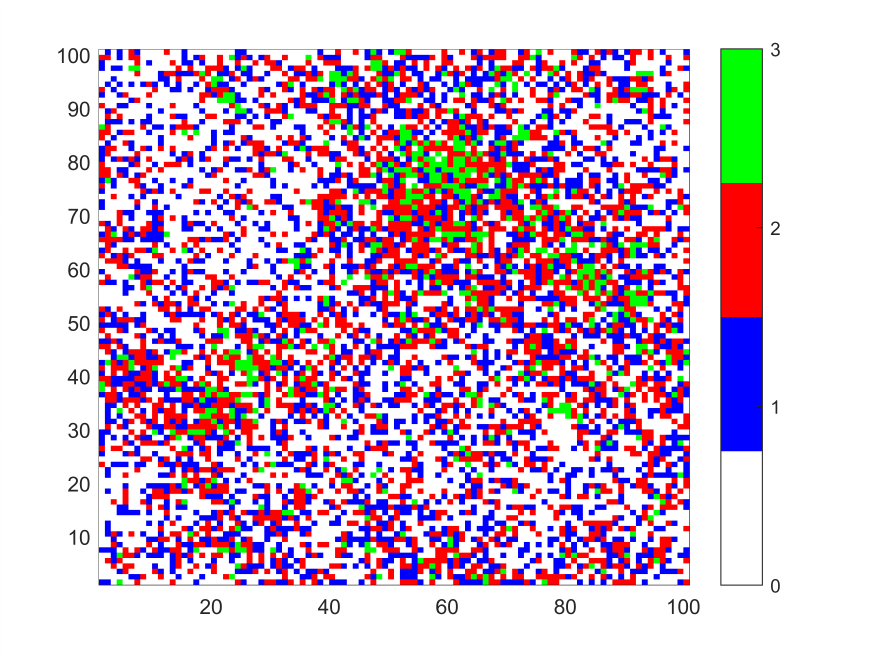}& 
	 			\includegraphics[scale=0.3]{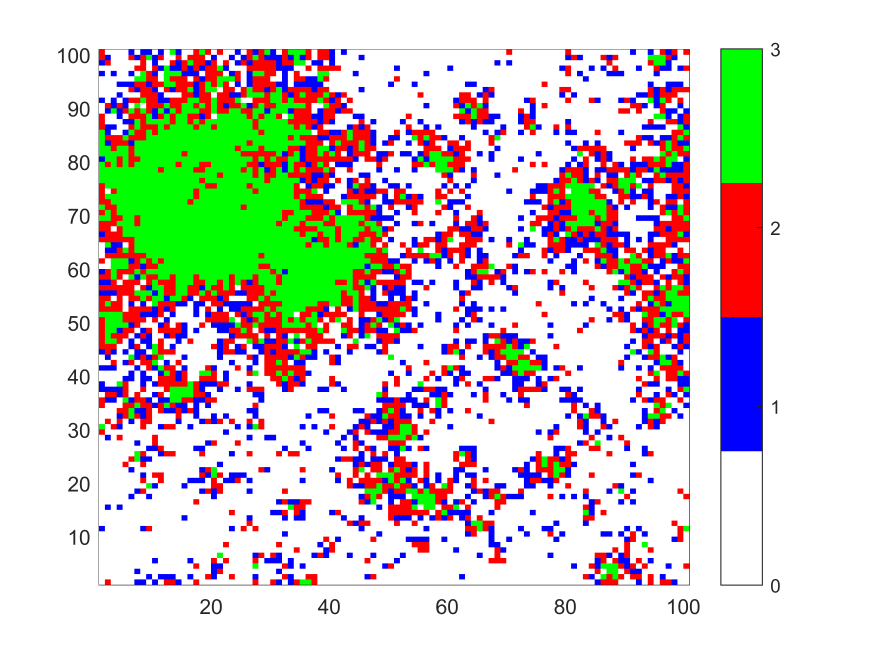}&
	 			\includegraphics[scale=0.3]{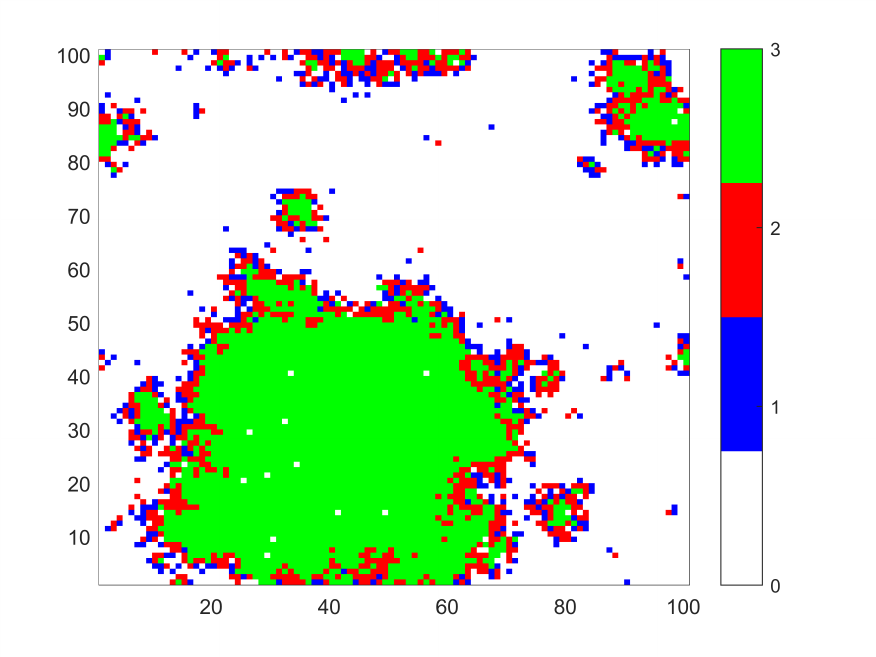}
			\end{tabular}
		\caption{Top row shows the $i(t)$ plots with $1\sigma$ interval for different values of $n$ and bottom row shows the spatial distribution of the disease spread at a intermediate time for the corresponding values of $n$.}\label{temp_sto}	
		\end{figure}
		To understand the reason behind the slower disease spread for higher $n$ we have plotted the time variation of $i(t)$ with the spatial distribution of the disease spread at a intermediate time for different values of $n$ (Fig.~\ref{temp_sto}). Here, the spatial distribution for a particular $n$ is chosen randomly from 50 samples. It is shown that the disease spread is more clustered for higher values of $n$. This is the reason for slower disease spread. Also, Fig.~\ref{n_var} shows that for $n=0$ the peak of $i(t)$ is lower than the $n=0.5$. The reason behind this is the variance $\sigma_{d}^{2}$. For this figure, $n=0.5$ have the highest $\sigma_{d}^{2}$ value amongst all $n$ values which have been used for this plot (Fig.~\ref{n_var}).
	
\section{Comparison with the continuous SEIR model}\label{cont_comp_sec}
In this section, we have compared our model with the continuous SEIR model. The model is described by the following equations:
\begin{equation}
	\frac{ds}{dt}=-\beta si
\end{equation}
\begin{equation}
	\frac{de}{dt}=\beta si-\sigma e
\end{equation}
\begin{equation}
	\frac{di}{dt}=\sigma e-\gamma i
\end{equation}
\begin{equation}
	\frac{dr}{dt}=\gamma i
\end{equation}

$\beta$= Infection rate.

$\sigma$= Latency period.

$\gamma$= Infectious period.

In this model, $\beta$, $\sigma$ and $\gamma$ are the free parameters. By definition we have assumed  $\sigma =\frac{1}{\tau_{\scriptscriptstyle{\text{I}}}}=0.125~day^{-1}$ and $\gamma = \frac{1}{\tau_{\scriptscriptstyle{\text{R}}}}=0.0556~day^{-1}$. We have generated various simulated data from the stochastic model for different $q$ values with $n$= 0.0, 0.5, 1.0, 1.5, 2.0, 2.5 and 3.0. Now for a particular $q$ value, we have fitted the continuous model on the simulated data corresponding to these $n$ values to find the best fit result for this particular $q$. To achieve this,  the sum of squared errors ($SSE$) of $i(t)$ has been minimized by varying the free parameter $\beta$ using an optimization algorithm, pattern search (PS).
\begin{equation}
	SSE=\sum_{k}\left(i_{k}-\hat{i_{k}}\right)^{2}
\end{equation}

$i$= Fraction of the infectious cases of the simulated data from our model.

$\hat{i}$= Fraction of the infectious cases from the continuous SEIR model.

	\begin{figure}[H]
	\centering
	\begin{tabular}{c c c c}
		\includegraphics[scale=0.3]{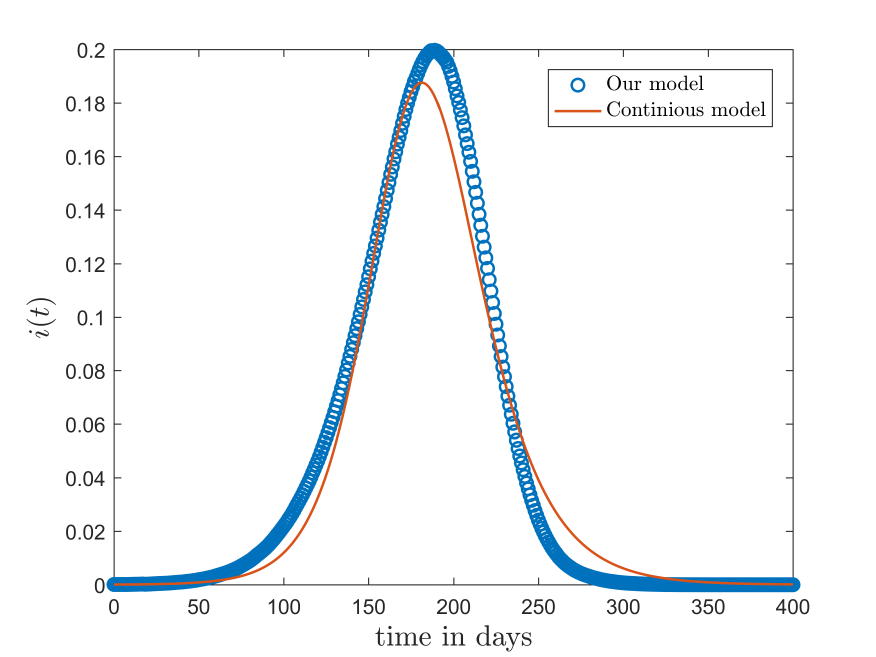}& 
		\includegraphics[scale=0.3]{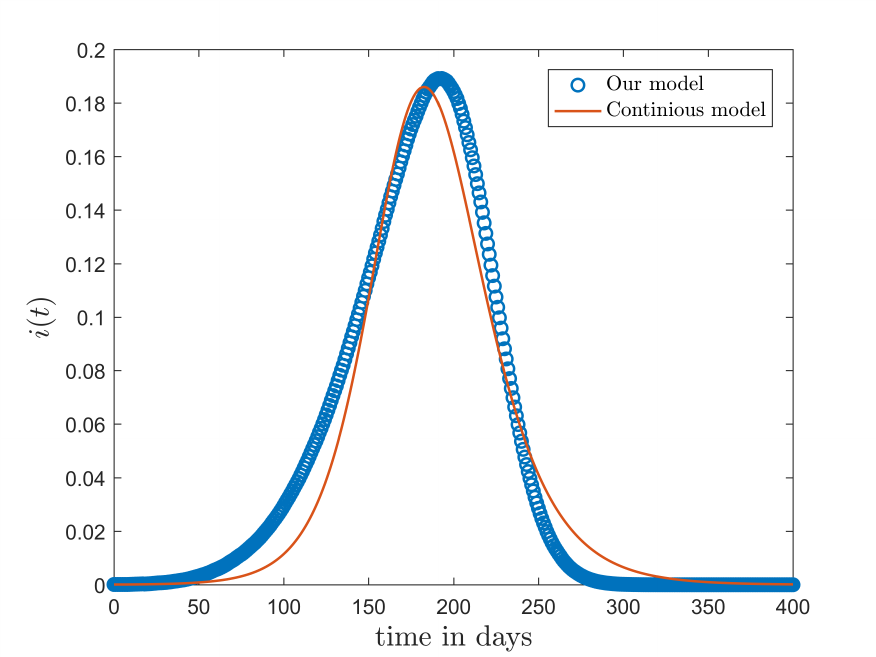}& 
		\includegraphics[scale=0.3]{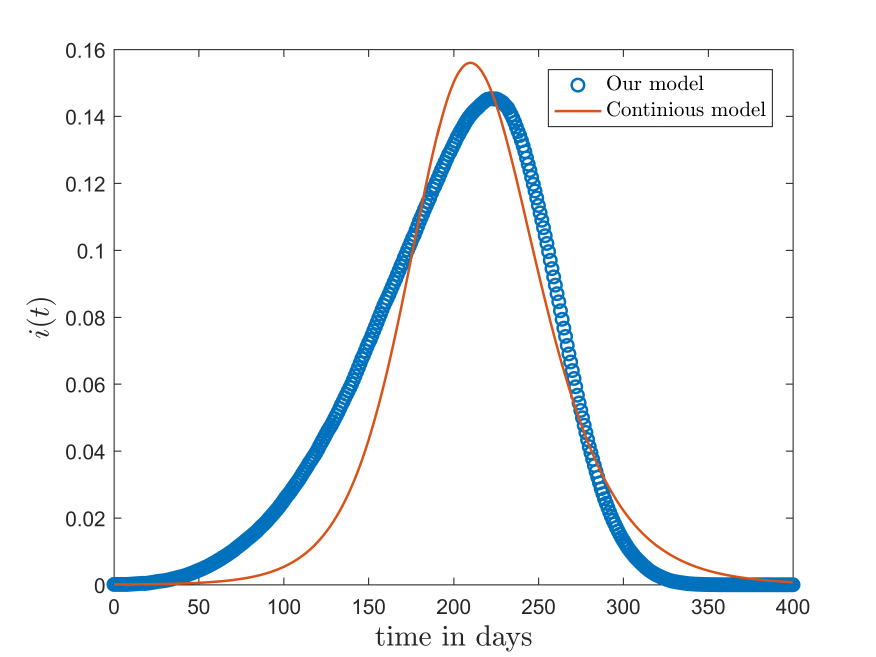}&
		\includegraphics[scale=0.3]{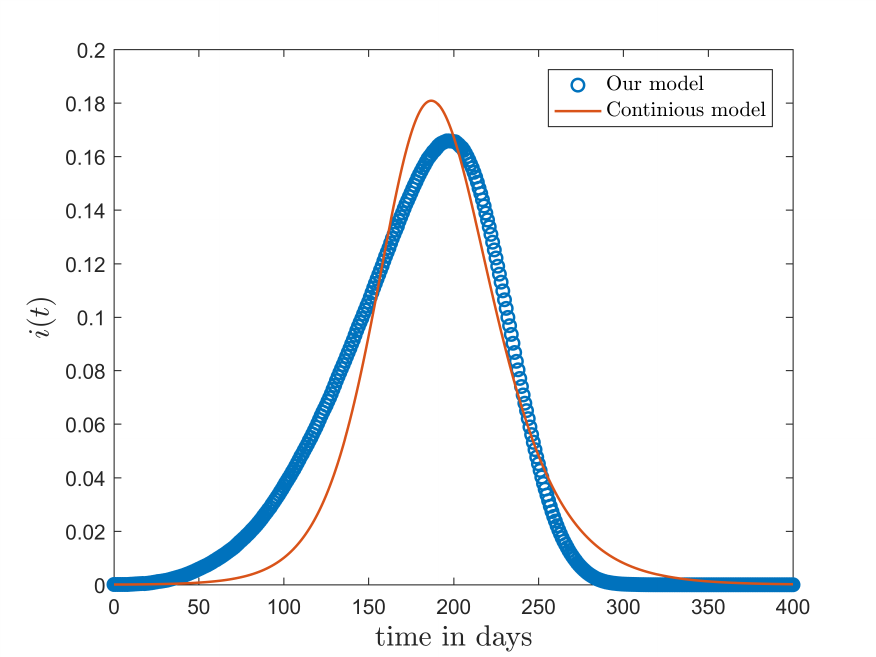}
	\end{tabular}
	\caption{Best fit curves of the continuous model to stochastic data for $q=0.2$, $q=0.3$, $q=0.4$, and $q=0.5$ (left to right order).}\label{cont_v_sto}	
\end{figure}

\begin{table}[H]
	\centering
	\begin{tabular}{|c|c|c|c|}
		\hline
		$q$ & $n$ & $\beta$ & $SSE$\\
		\hline
		0.2 &	2.0	&	0.1554	&	3.90E-02	\\	
		\hline
		\hline
		0.3	&	2.5	&	0.1544	&	5.73E-02	\\
		\hline
		\hline
		0.4	&	3.0	&	0.1367	&	7.94E-02	\\
		\hline
		\hline
		0.5	&	3.0	&	0.1512	&	8.48E-02	\\	
		\hline
		\hline
		0.6	&	3.0	&	0.1677	&	9.11E-02	\\	
		\hline                     
	\end{tabular}
	\caption{{Pair of $q$ and $n$ values for which continuous model fit well on the CA model data. The optimized values of $\beta$ of the continuous model and corresponding $SSE$ values are also shown here.} \label{SSE_tab}}
\end{table}
Tab.~\ref{SSE_tab} shows different $q$ values and corresponding $n$ values for which continuous model fit well on our CA model data. Also, this table shows the optimized $\beta$ values of the continuous model and $SSE$ values of the respective cases. In Fig.\ref{cont_v_sto} shows the best fit results which are given in Tab.~\ref{SSE_tab}. From these results we can say that continuous model fits better on our model for higher $n$.                                  
\section{Comparison with other neighbourhood conditions}\label{ngh_comp_sec}
In our model, we have introduced a different neighbourhood condition (refer to Sec.~\ref{sec2} for a detailed discussion) than those favored in the current literature. In this context, we discuss and compare our neighbourhood condition with other neighbourhood conditions in this section.
\subsection{Moore's neighbourhood condition}\label{mr_comp}
\begin{figure}[H]
	\centering
	\includegraphics[scale=0.5]{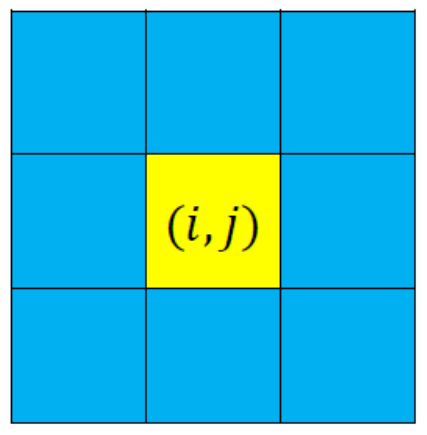}
	\caption{{This figure shows Moore's neighbourhood condition where the blue cells represent nearest neighbourhood of the cell, $(i,j)$.}\label{moore_ngh}}	
\end{figure}
\begin{figure}[H]
	\centering
	\includegraphics[scale=0.65]{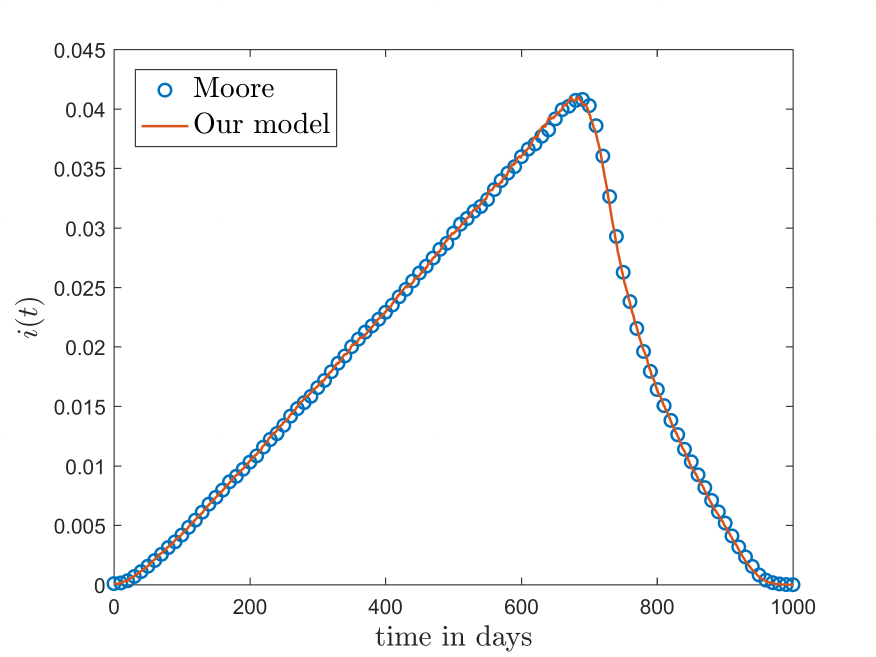}
	\caption{Comparison of the model with Moore's neighbourhood condition and our model for $n=10$.}\label{moore_comp}
\end{figure}

In Moore's neighbourhood condition any cell can interact only with those neighbours which belongs to the first layer \cite{Moore}. In our model when $n>3$, $\langle d\rangle\approx 1$ and $\sigma_{d}^{2}$ decreases with increasing $n$. Thus our model with degree exponent $n\gg 3$ will give a similar result like a model with Moore's neighbourhood condition. Fig.\ref{moore_comp} shows the comparison between Moore's neighbourhood condition and our model for $n=10$.

\subsection{$r$-neighbourhood}\label{dn_comp}
\begin{figure}[H]
	\centering
	\includegraphics[scale=0.65]{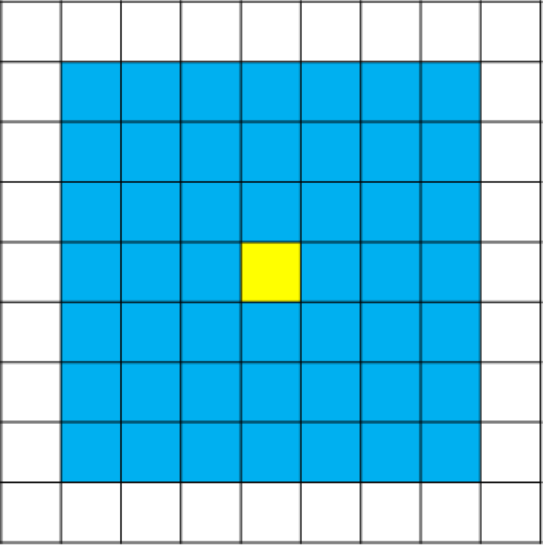}
	\caption{$r$-neighbourhood condition for $r=3$. Here, blue cells represent nearest neighbourhood of the cell, $(i,j)$.}\label{ext_nghbd}
\end{figure}

Another one of the most popular neighbourhood conditions, which has been used in many CA models to study with the COVID-19 pandemic, is the $r$-neighbourhood condition \cite{covid_social_isol,covid_vac_lockdwn, covid_GA_1}. In this neighbourhood condition, a cell can interact any other cell in a radius of $r$. Thus the probability of interaction of a cell with any other cells upto the $r$th layer is assumed to be equal. Fig.~\ref{ext_nghbd} shows this neighbourhood condition for $r=3$.
Thus the interaction probability of a cell to another cell of layer $\ell$, where $\ell\leq r$ is 
\begin{equation}
	p_{\scriptscriptstyle{\text{int}}}(\ell)=\frac{8\ell}{n_{r}^{2}-1},\qquad\ell\leq r.
\end{equation}
Here, $8\ell$ is the total number of cells in layer $\ell$ and $n_{r}$ is the total number of cells at one side of the $r$th layer which have value $n_{r}=2r+1$. Thus for a $r$-neighbourhood model, the probability of interaction varies linearly with layer number ($\ell$) when $\ell\leq r$ and goes to zero for $\ell>r$. Hence if we take the value of the degree exponent as, $n=-1$ for $\ell\leq r$ and $n\gg 3$ when $\ell>r$ in our model (referred to Eq.~\ref{pint_rel}) then it gives the similar result as for the $r$-neighbourhood model.

Thus in our neighbourhood condition, we have introduced a discontinuity at the layer $\ell=r$ for defining our probability of interaction to reproduce the results of the $r$-neighbourhood condition. Thus,  ($p_{\scriptscriptstyle{\text{int}}}$) is mathematically expressed as, 
\begin{equation}
	p_{\scriptscriptstyle{\text{int}}}(\ell)\propto \frac{1}{\ell^{n}},~n=-1\qquad \ell\leq d.
\end{equation}
\begin{equation}
	p_{\scriptscriptstyle{\text{int}}}(\ell)\propto \frac{1}{\ell^{n}},~n\gg 3 \qquad \ell>d.
\end{equation}
	\begin{figure}[H]
	\begin{subfigure}{.33\textwidth}
		\centering
		\includegraphics[scale=0.4]{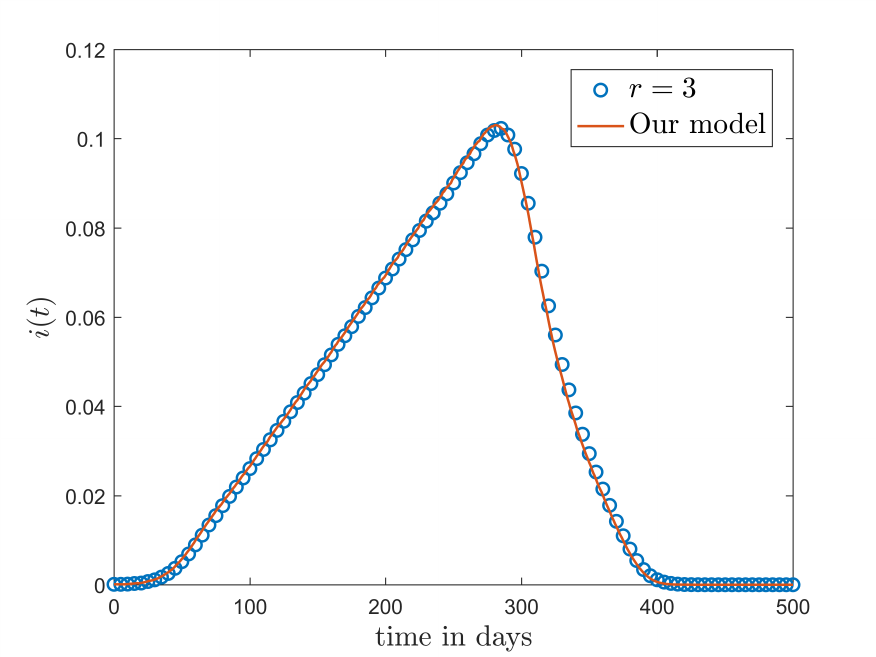}
		\caption{}
		\label{ext_1}
	\end{subfigure}
	\begin{subfigure}{.33\textwidth}
		\centering
		\includegraphics[scale=0.4]{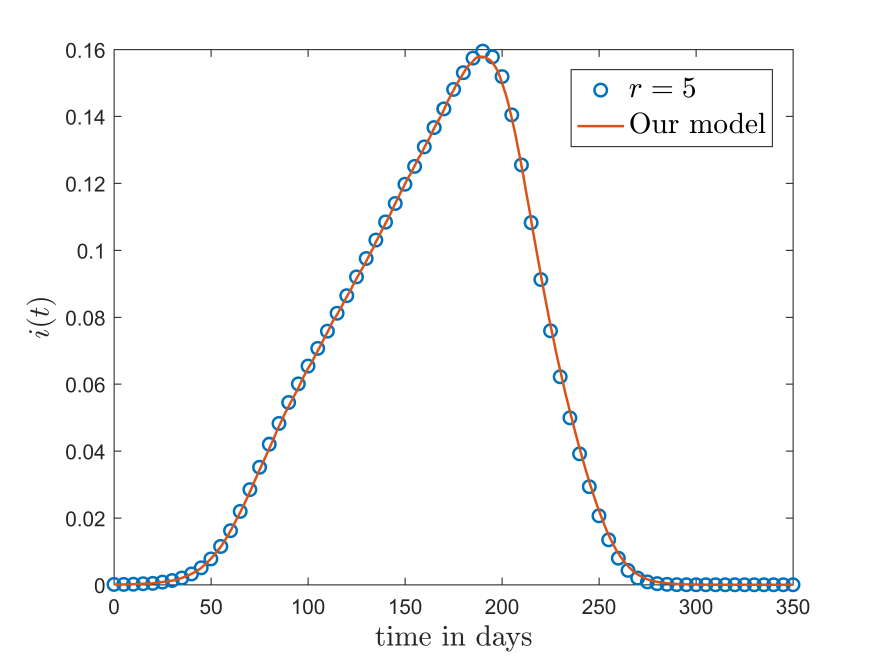}
		\caption{}
		\label{ext_2}
	\end{subfigure}
	\begin{subfigure}{.33\textwidth}
		\centering
		\includegraphics[scale=0.4]{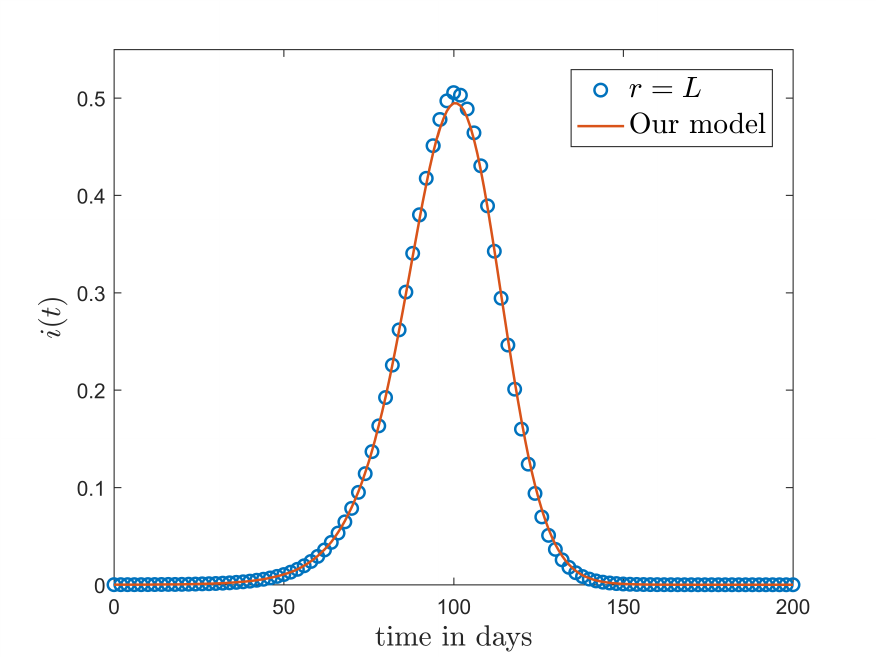}
		\caption{}
		\label{random_comp}
	\end{subfigure}
	\caption{Comparison of different $r$-neighbourhood models with our model. (a) $r=3$, (b) $r=5$ and (c) $r=L$, $L$ is the total number of layers in a $N\times N$ lattice.}\label{extended_comp}	
\end{figure}

Fig.~\ref{ext_1} and \ref{ext_2} shows the comparison between the $r$-neighbourhood model and our model for $r=3$ and $r=5$ respectively. Also for a $N\times N$ lattice, if we choose $r=L$, where $L$ denotes the total number of layers in the lattice and given by Eq.~\ref{L_rel} then one cell can interact with any other cell of the lattice. The comparison of the results for $r$-neighbourhood model with $r=L$ and our model is shown in Fig.~\ref{random_comp}.

\section{COVID-19 data and our model}\label{cov_fit_sec}
Here, we have discussed the results obtained from fitting of our model on multiple peaks of COVID-19 data for different countries. Here, we have considered a total of eight countries: Italy, France, Germany, Brazil, US, India, South Africa and Japan. The arrangement and number of peaks in daily data of new cases varies with respect to different countries. Also, the size of the peaks differ for different countries. Thus to fit the model in COVID-19 data we have made following assumptions:

\begin{enumerate}
	\item We have explored only those part of COVID-19 data where there has been a sharp increase of daily new cases and have ignored those parts where daily data for new cases remains low for a long time.
	\item Every peak of this disease spread was considered separately for fitting. When a particular peak other than the first peak is modeled, the effective susceptible population that has been considered includes both new susceptible people and the people who had been infected but have again become susceptible due to lack of sustainable immunity against COVID-19.
	\item Each peak is considered as a new disease spread in a population. It is like a fresh start. So, it does not have any relation with the disease spread previous to it. The peaks of COVID-19 for a particular country have different features and reasons of occurrence. The number of infected persons as well as the time span for these peaks are different. Also, there could be multiple reasons for their incidence like mutation in the SARS-COV-2 virus or decrement of social restrictions. CA models have been made with finite lattices and we need to fix its size throughout the study. However, the number of effective susceptible people (susceptible persons who take part in the disease dynamics) in these peaks are different due to the variation of the total number of infections between peaks. Thus it will be difficult to model all the peaks together by a single CA model. 
\end{enumerate}
We have taken data from the COVID-19 data repository of Johns Hopkins University Center for Systems Science and Engineering (JHU CSSE) \cite{JHU_data}. This repository contains time series data of confirmed cases and deaths for different countries. However, the time series data for recovered cases is discontinued at a very early stage. Our model needs removed cases (Recovered and dead both) data, which was not available in this database. Thus we have made further assumptions which are enlisted below:
\begin{enumerate}
	\item Number of deaths ($D(t)$) and removed cases ($R(t)$) at any time $t$ is related by,
	\begin{equation}\label{death_rel}
		D(t)=k(t)R(t)
	\end{equation}
	Here $k(t)$ is a function of time $t$ and has a value between 0 and 1.
	\item Number of deaths during the COVID-19 pandemic not only depended on the fatality of the virus but also on the condition of the medical infrastructure. New variants of SARS-COV-2 virus appeared at different times as well as the medical supplies and the healthcare facilities were not always the same during this entire COVID period. Thus it is a better proposal to keep $k$ dynamic with time. However, to preserve the simplicity of our model, we have assumed that during a particular peak in a country, $k$ is constant.
\end{enumerate}

Here the model is fitted to the confirmed cases data. Since we have considered the peaks separately, the confirmed cases data for a particular peak required proper scaling before fitting. We suppose that any one chosen peak starts at time $T_{1}$ and ends at $T_{2}$. Thus to scale the confirmed cases data in the range $T_{1}$ to $T_{2}$, we have to subtract the confirmed cases data in this range with the confirmed cases at time $T_{1}-1$. So the scaled confirmed cases for a particular peak $P$ can be written as,
\begin{equation}\label{scale1}
	I_{C}^{P}(t)=I_{C}(t)-I_{C}(T_{1}-1)\qquad T_{1}\leq t\leq T_{2}
\end{equation}
Also the confirmed cases for peak $P$ ($I_{C}^{P}(t)$) has been normalized as,
\begin{equation}
	i_{C}^{P}(t)=\frac{I_{C}^{P}(t)}{I_{C}^{P}(T_{2})}\qquad T_{1}\leq t\leq T_{2}
\end{equation}
The time scale of $i_{C}^{P}(t)$ has also been changed from $T_{1}\leq t\leq T_{2}$ to $0\leq t\leq T_{2}-T_{1}$. 

These scaled and normalized confirmed cases, $i_{C}(t)$ in $0\leq t\leq T$ (omitting the superscript and assuming $T_{2}-T_{1}=T$) is fitted with the normalized total infected cases, $i_{tot}(t)$ in the same timescale. Here the total infected cases (from our model) is normalized as,

\begin{equation}
	i_{tot}(t)=\frac{I_{tot}(t)}{I_{tot}(T)}\qquad 0\leq t\leq T
\end{equation}

We have used Genetic Algorithm (GA) to optimize the following Sum of Squared Errors (SSE) as given in Eq.~\ref{SSE_GA} to fit our model to the data.
\begin{equation}\label{SSE_GA}
	SSE=\sum_{t=0}^{T}(i_{C}(t)-i_{tot}(t))^2
\end{equation} 
Here, 

$i_{C}(t)$= Fraction of confirmed cases for a particular peak.

$i_{tot}(t)$= Fraction of total cases for that peak simulated from the model.

For fitting, we have assumed a squared lattice of size $N\times N=101\times 101$. Our model has four free parameters ($q, n, \tau{\scriptscriptstyle{\text{I}}}, \tau_{\scriptscriptstyle{\text{R}}}$) and needs three initial conditions ($E(0), I(0), R(0)$). We have assumed that, $E(0)=0$, $R(0)=0$ and taken $I(0)$ as a free parameter. Thus, Eq.~\ref{SSE_GA} is optimized with five free parameters $q, n, \tau_{\scriptscriptstyle{\text{I}}}, \tau_{\scriptscriptstyle{\text{R}}}$ and $I(0)$.

 To compare the model results with the data representing the number of deaths, we need a slightly different scaling than the one which is described in Eq.~\ref{scale1}. According to our model, an infectious person will be moved to the removed compartment after $\tau_{\scriptscriptstyle{\text{R}}}$ days.Thus, for $0\leq t<\tau_{\scriptscriptstyle{\text{R}}}$, the removed cases, $R(t)$ as well as deaths, $D(t)$ is zero. Suppose $D^{P}(t)$ denotes the number of deaths at any time $t$ for a particular peak $P$ which is ranged from $T_{1}\leq t\leq T_{2}$. We have assumed each peak separately, however the number of deaths will not be zero in the range $T_{1}\leq t<T_{1}+\tau_{\scriptscriptstyle{\text{R}}}$. Number of deaths in this range depends on the number of infectious cases (or, active cases, both are same in our model) at $t<T_{1}$, which are not zero generally. Thus to compare our model to the data representing deaths, we have to omit number of deaths in the range $T_{1}\leq t<T_{1}+\tau_{\scriptscriptstyle{\text{R}}}$. Thus the proper scaling for the deceased cases is,
\begin{equation}
	D^{P}(t)=D(t)-D(T_{1}+\tau_{\scriptscriptstyle{\text{R}}}-1)\qquad T_{1}+\tau_{\scriptscriptstyle{\text{R}}}\leq t\leq T_{2}
\end{equation}
If we re-scale the time axis, then the above relation will have a form,
\begin{equation}\label{death_delay}
	D^{P}(t)=D(t)-D(\tau_{\scriptscriptstyle{\text{R}}}-1)\qquad \tau_{\scriptscriptstyle{\text{R}}}\leq t\leq T
\end{equation}

Thus, in this paper, we have fitted our model to the data by considering each peak separately by minimizing Eq.~\ref{SSE_GA}. We have used GA to optimize  Eq.~\ref{SSE_GA}. We have run GA multiple times to get best fit parameter values. Then we have re-checked these parameters values by comparing the variation of deaths (simulated) with the deceased cases data. 

\begin{figure}[H]
	\centering
	\begin{tabular}{c c c}
		\includegraphics[scale=0.4]{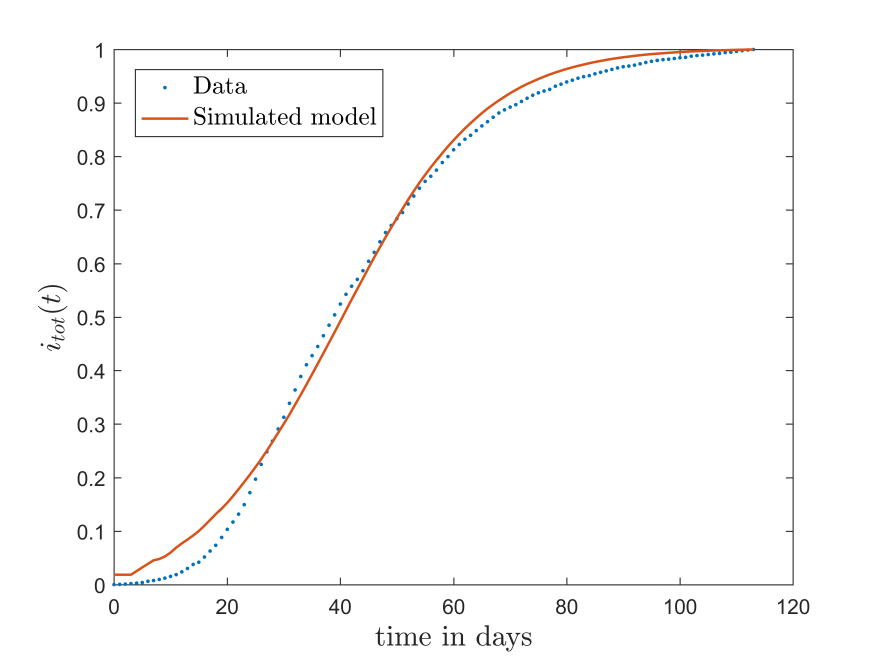}&
		\includegraphics[scale=0.4]{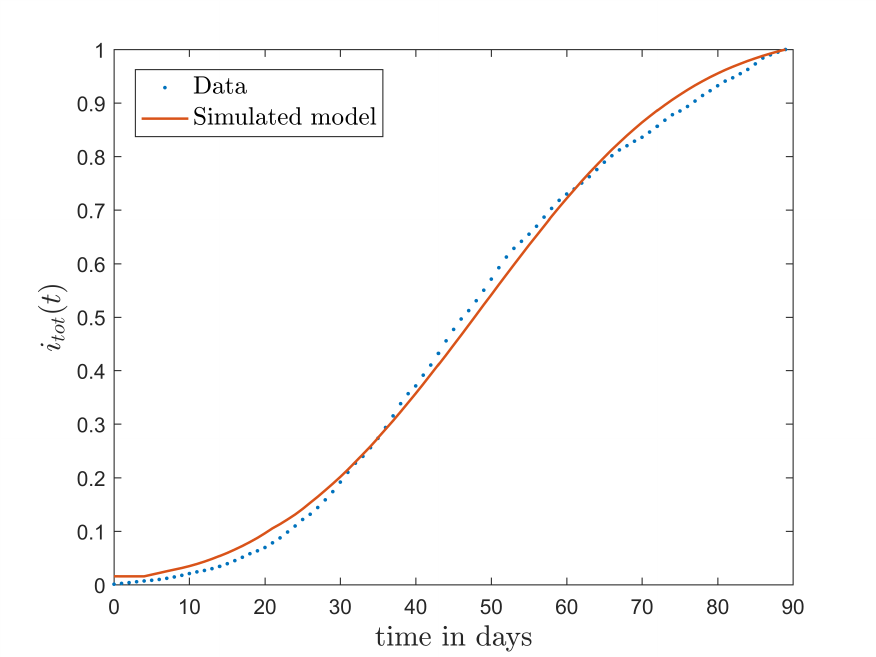}& 
		\includegraphics[scale=0.4]{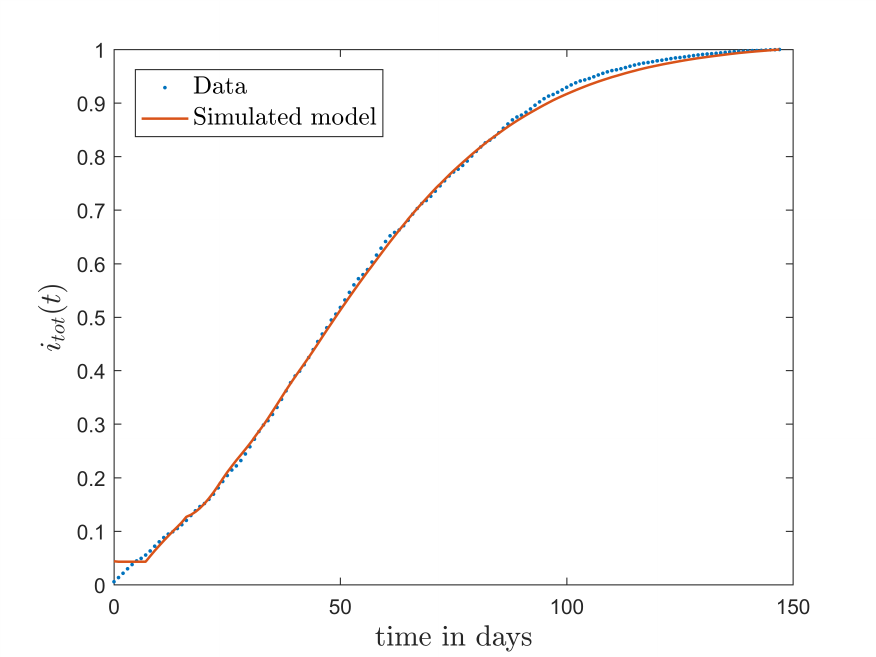}
	\end{tabular}
	\centering
	\begin{tabular}{c c c}
		\includegraphics[scale=0.4]{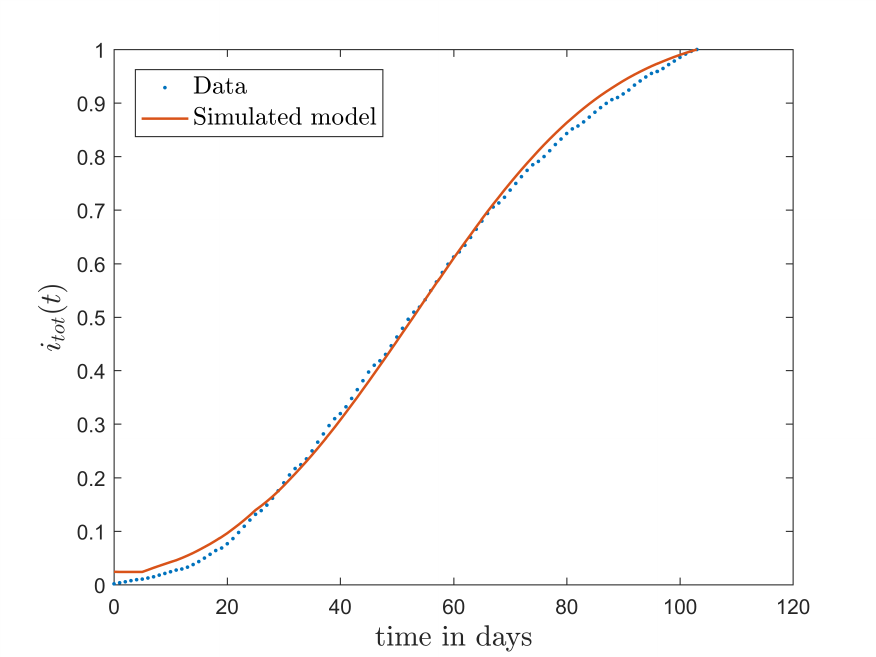}&
		\includegraphics[scale=0.4]{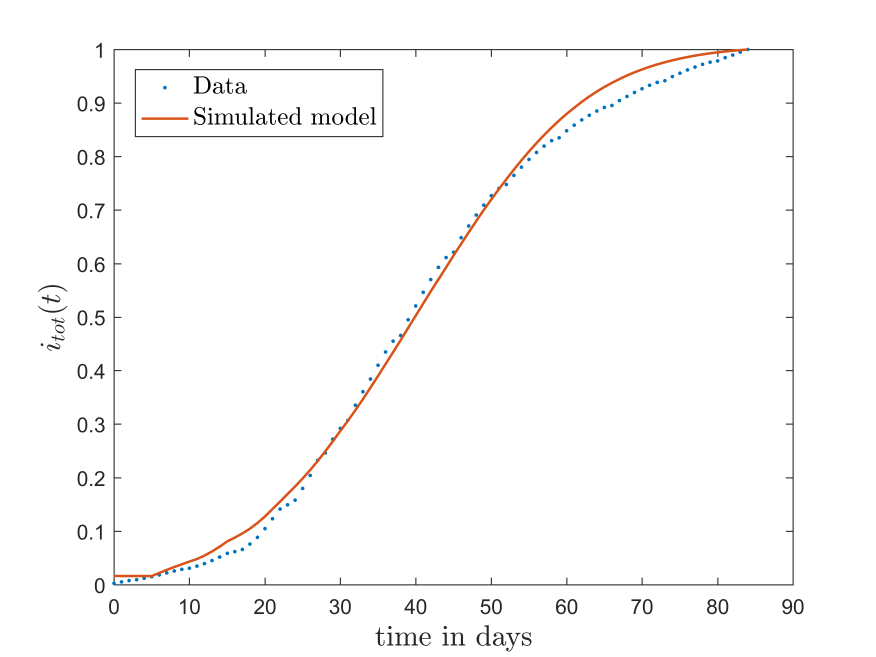}&
		\includegraphics[scale=0.4]{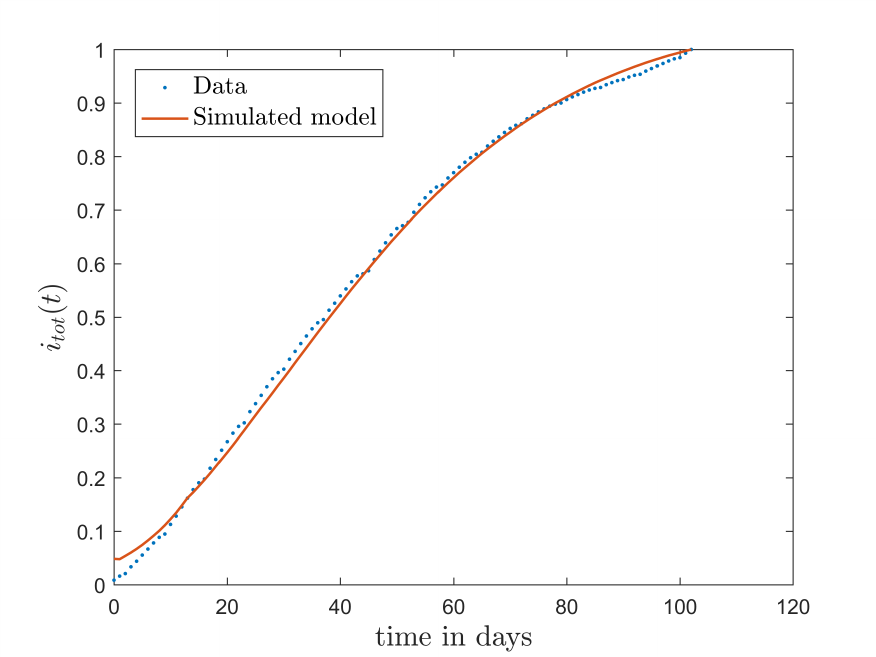}
	\end{tabular}
	\caption{Model fitting to the confirmed cases data of Italy for five different peaks. Peaks are counted from left to right and top to bottom.}\label{Italy_sep_itot}	
\end{figure}

\begin{figure}[H]
	\centering
	\begin{tabular}{c c c}
		\includegraphics[scale=0.4]{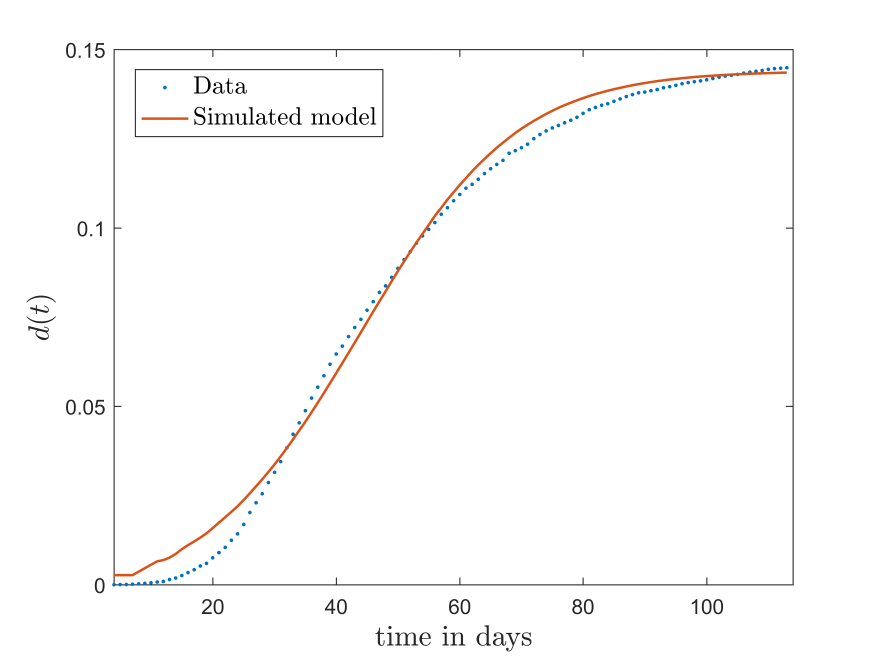}&
		\includegraphics[scale=0.4]{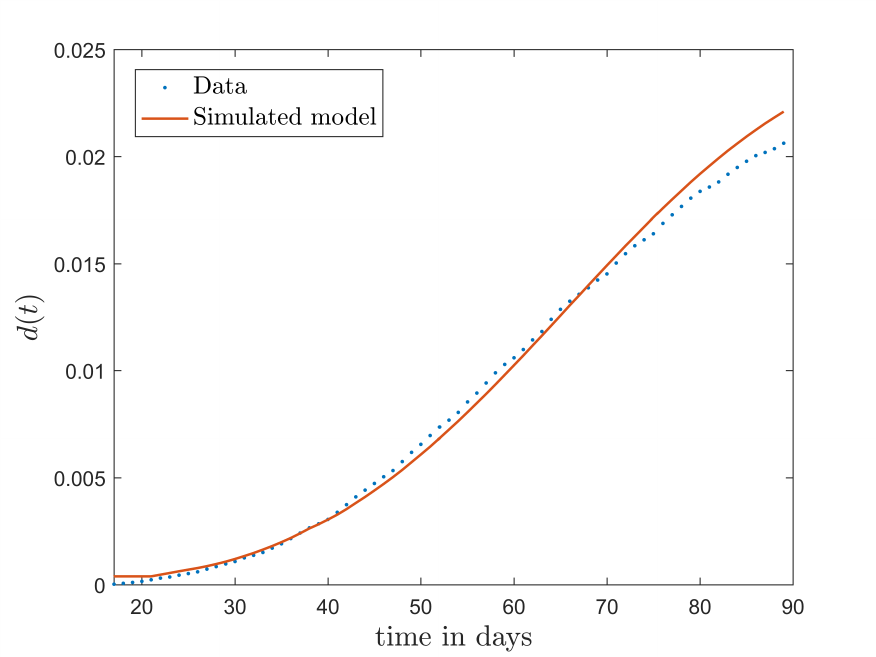}& 
		\includegraphics[scale=0.4]{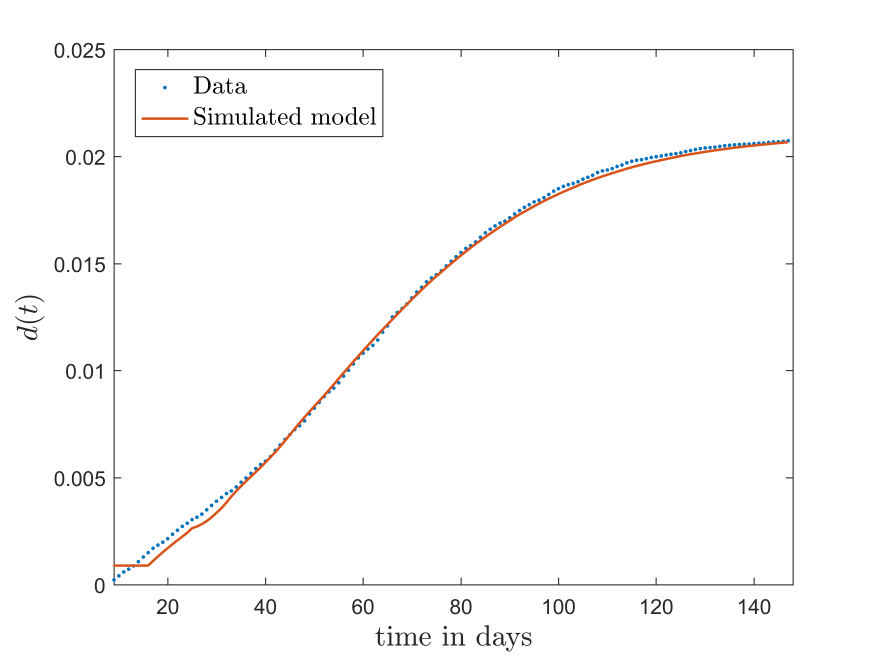} 
	\end{tabular}
	\centering
	\begin{tabular}{c c c}
		\includegraphics[scale=0.4]{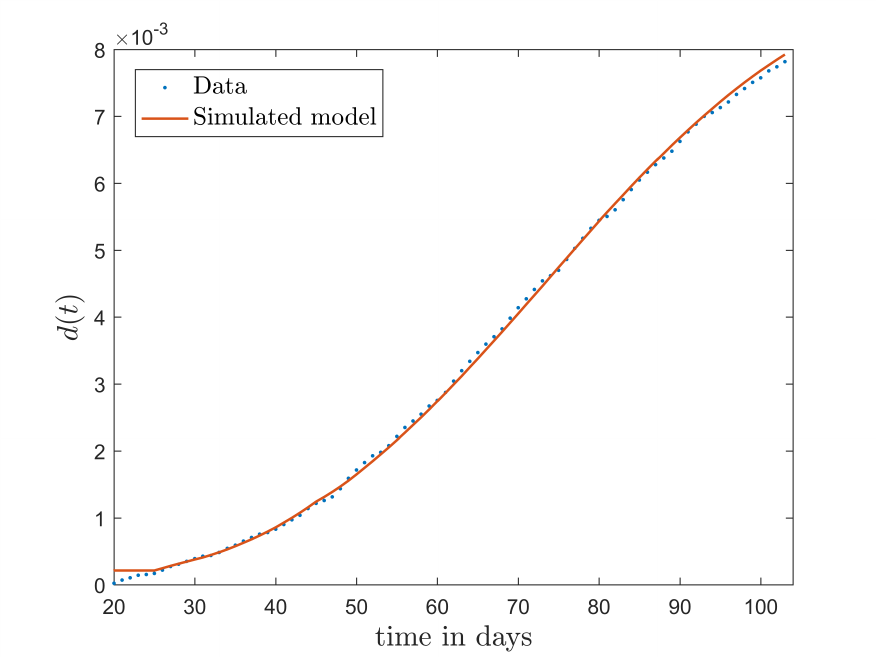}&
		\includegraphics[scale=0.4]{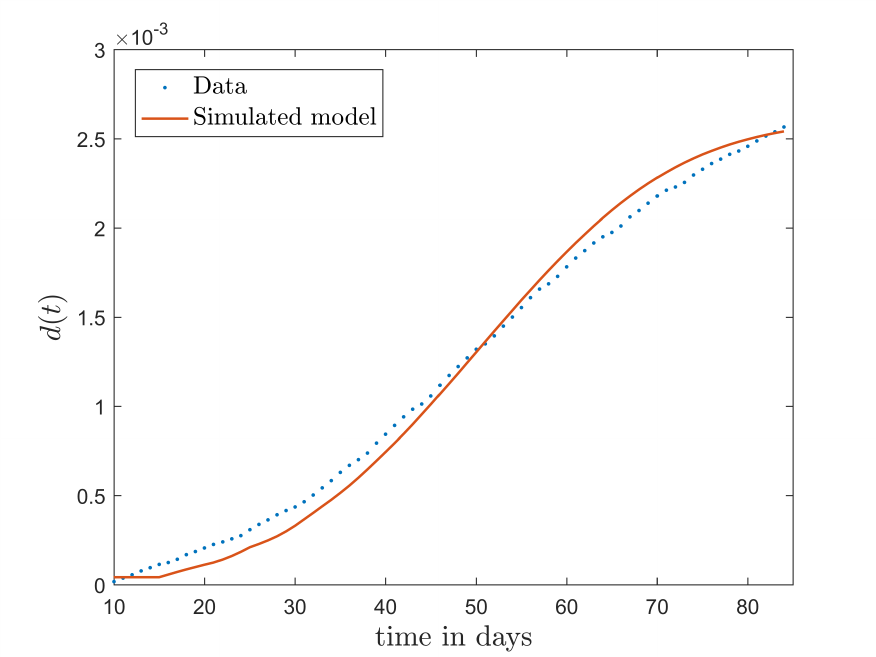}&
		\includegraphics[scale=0.4]{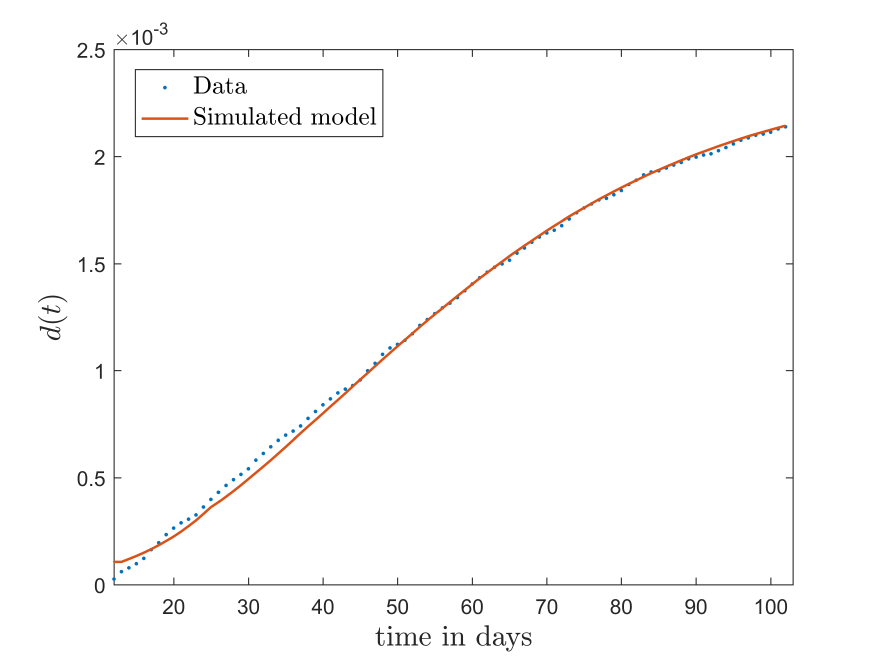}
	\end{tabular}
	\caption{Model fitting to the death data of Italy for five different peaks. Peaks are counted from left to right and top to bottom.}\label{Italy_sep_d}
\end{figure}

In Fig.~\ref{Italy_sep_itot}, we show our model is fitted to the confirmed cases data of six peaks (with proper scaling) of Italy. Also, we have used best fit values of the parameters to simulate the variations of the fraction of death cases for these peaks and have compared them with the scaled deceased cases data of Italy in Fig.~\ref{Italy_sep_d}.

\begin{figure}[H]
	\begin{subfigure}{.5\textwidth}
		\centering
		\includegraphics[scale=0.65]{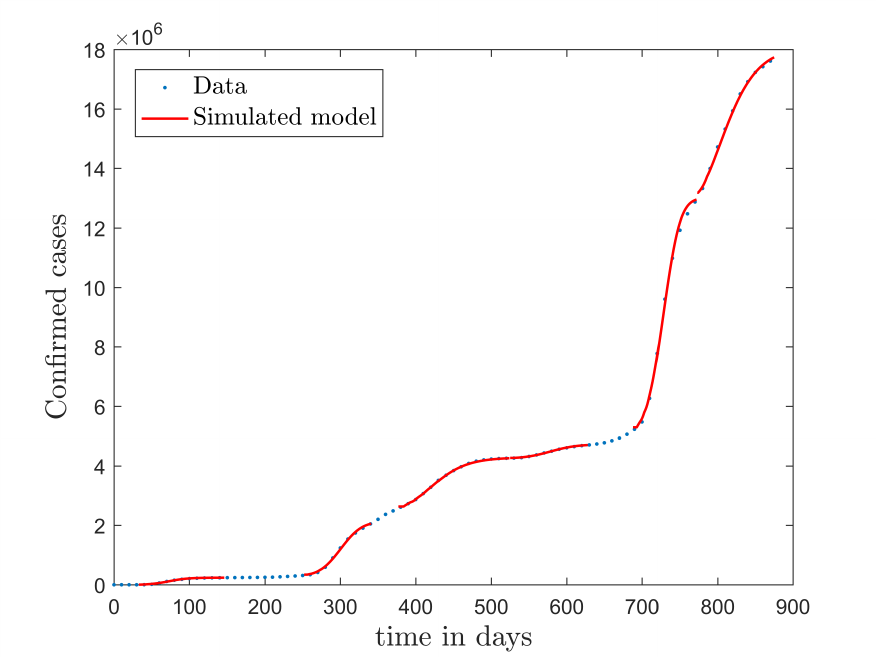}
		\caption{}
		\label{It_it}
	\end{subfigure}
	\begin{subfigure}{.5\textwidth}
		\centering
		\includegraphics[scale=0.65]{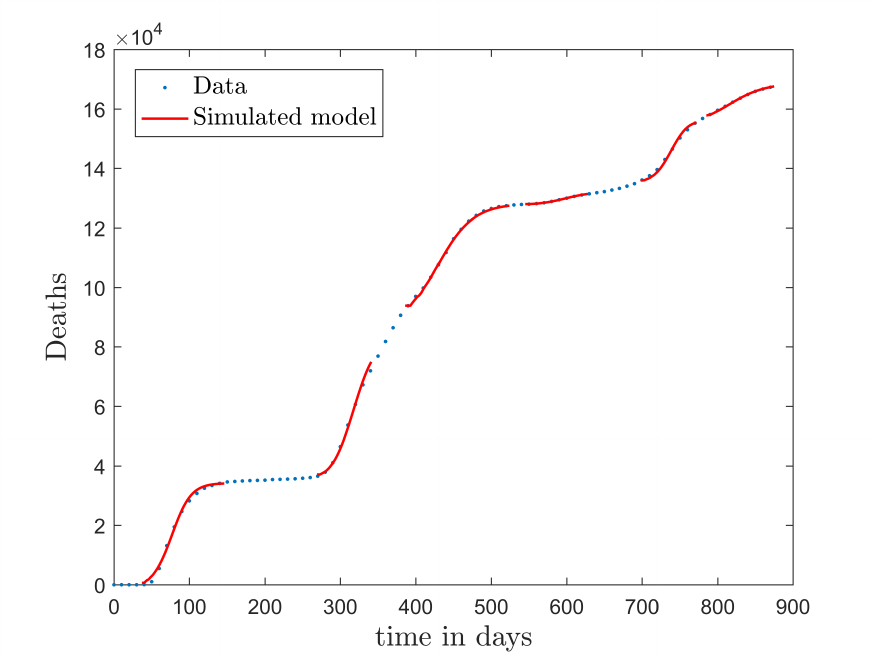}
		\caption{}
		\label{It_d}
	\end{subfigure}
	\caption{All fitted plots for five different peaks are combined and plotted  against the whole data of Italy with proper re-scaling.}\label{Italy_all}	
\end{figure}	

\begin{figure}[H]
	\centering
	\begin{tabular}{c c}
		\includegraphics[scale=0.55]{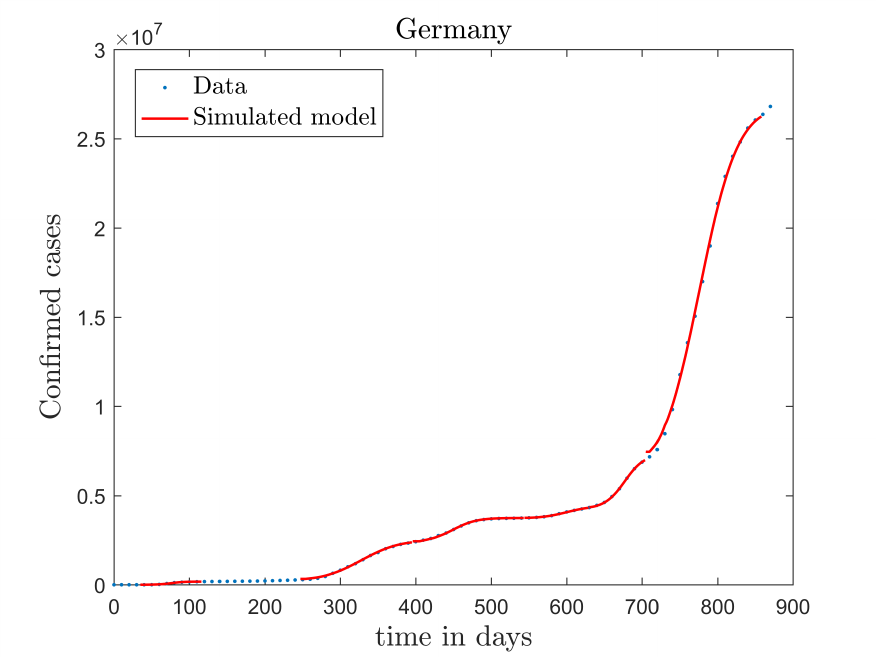}& 
		\includegraphics[scale=0.55]{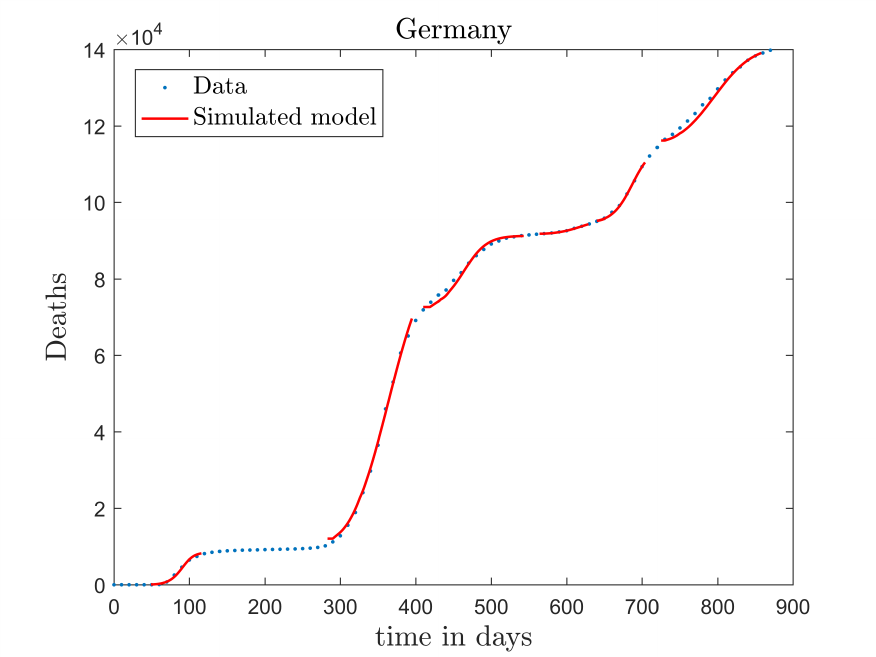}\\
		\includegraphics[scale=0.55]{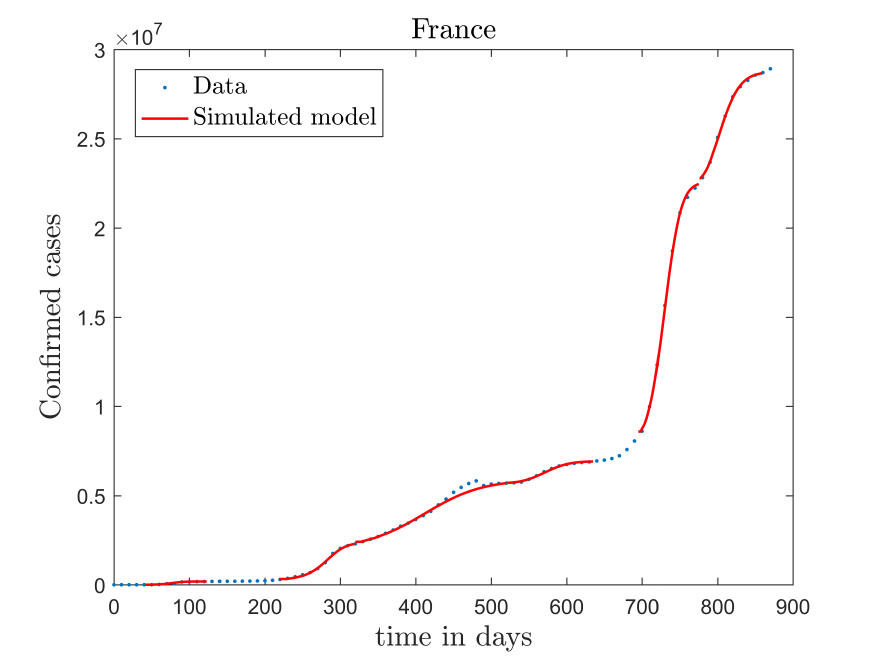}& 
		\includegraphics[scale=0.55]{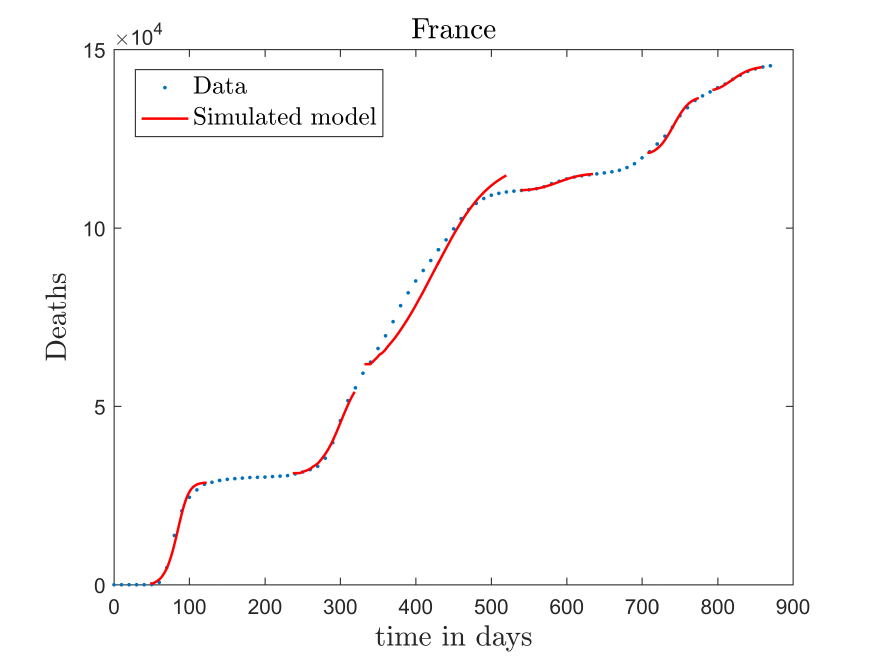}\\
		\includegraphics[scale=0.55]{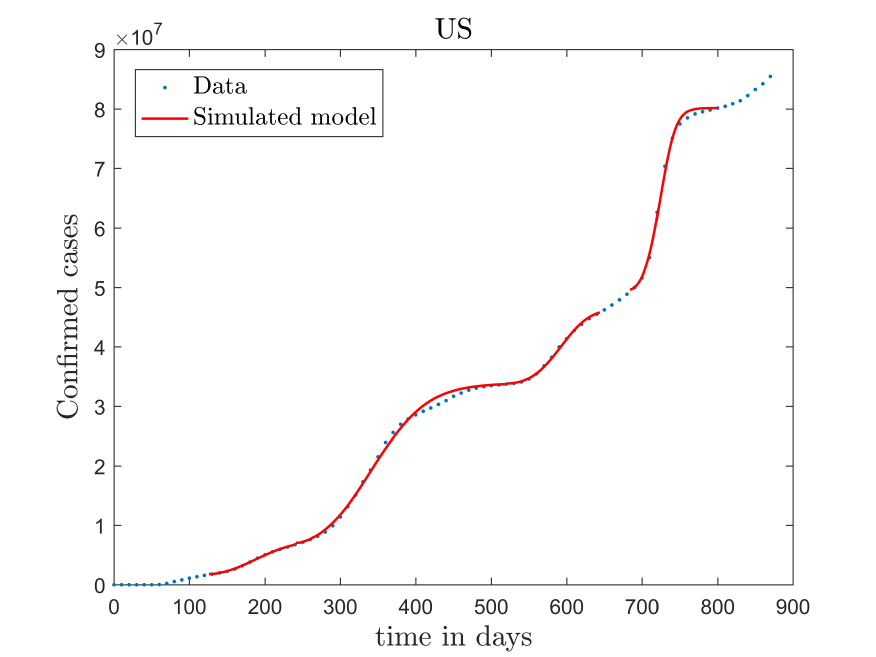}& 
		\includegraphics[scale=0.55]{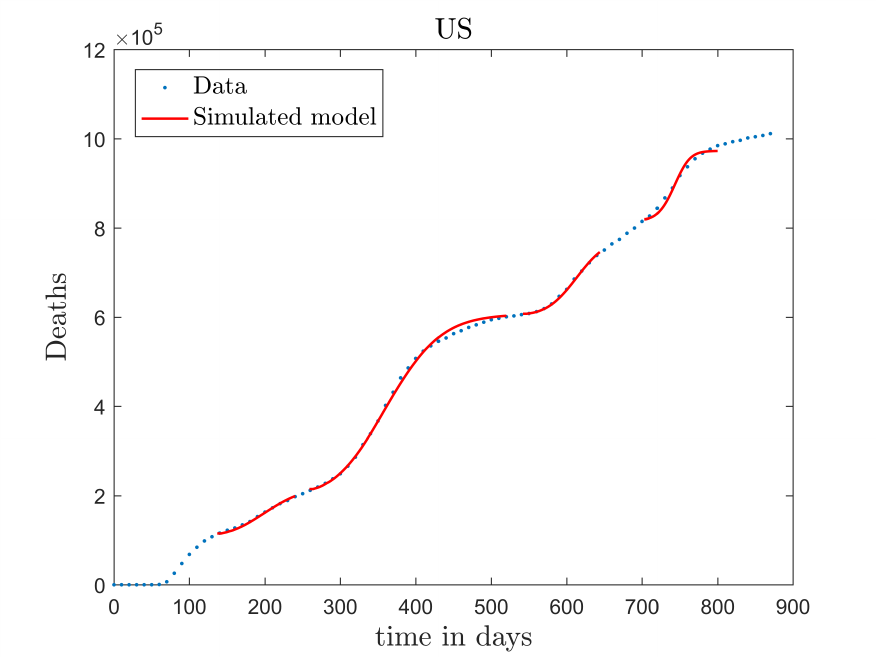}\\
		\includegraphics[scale=0.55]{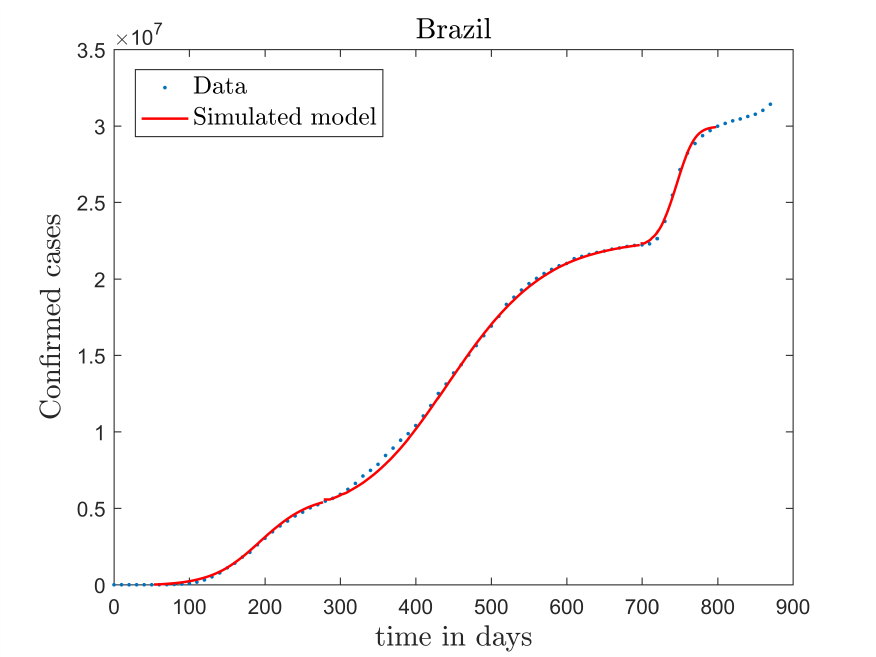}& 
		\includegraphics[scale=0.55]{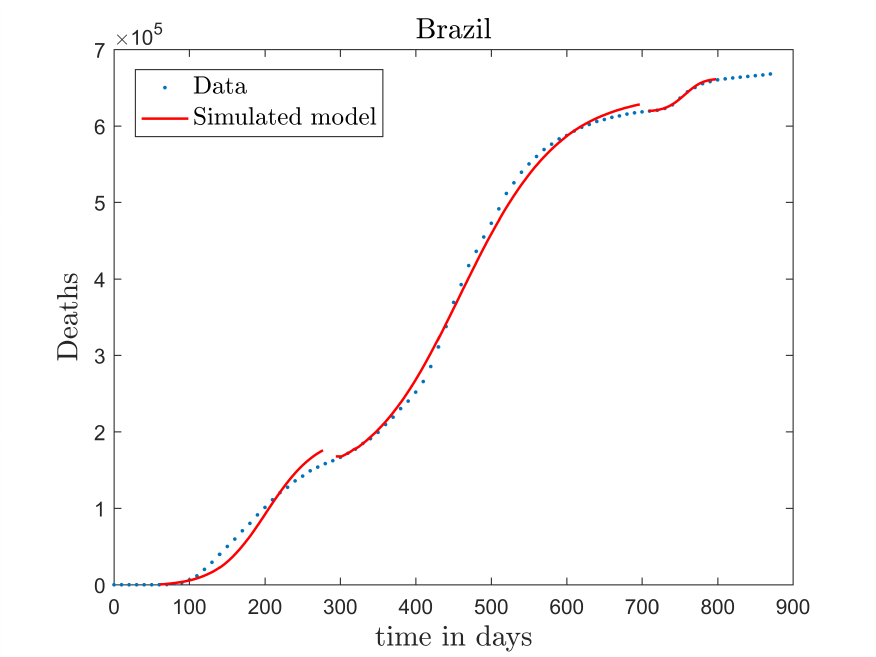}
	\end{tabular}
\end{figure}

\begin{figure}[H]
	\centering
	\begin{tabular}{c c}
		\includegraphics[scale=0.55]{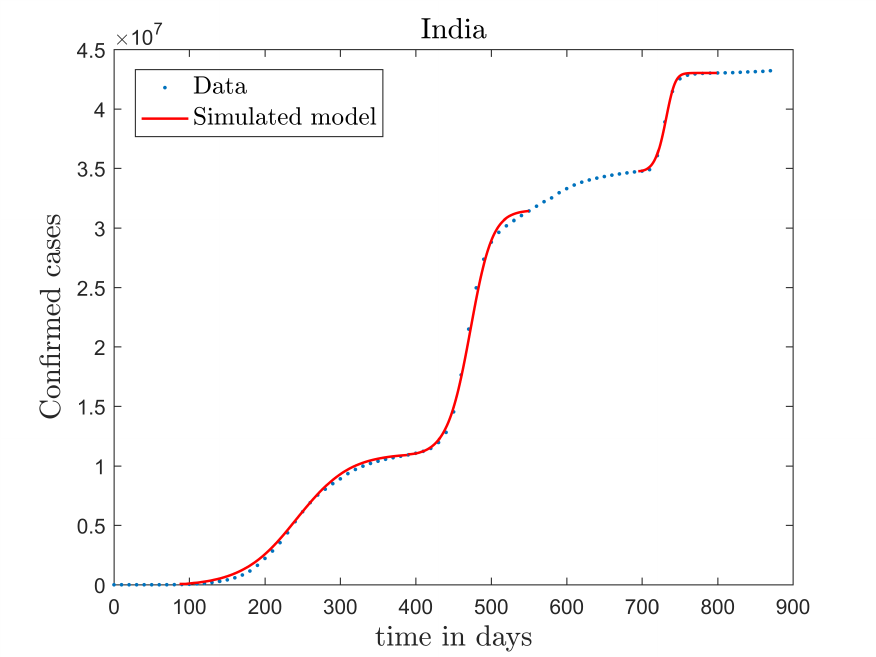}& 
		\includegraphics[scale=0.55]{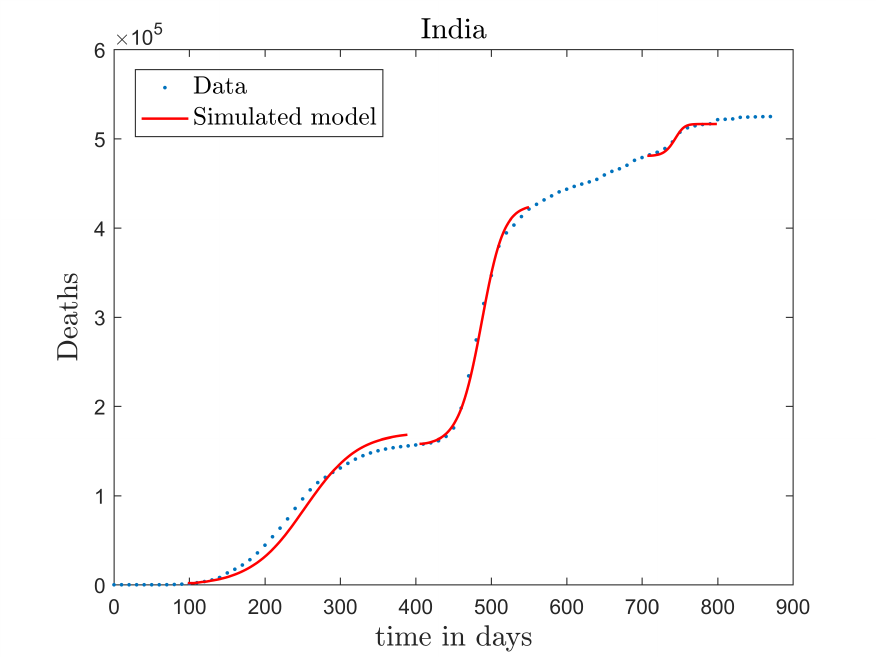}\\
		\includegraphics[scale=0.55]{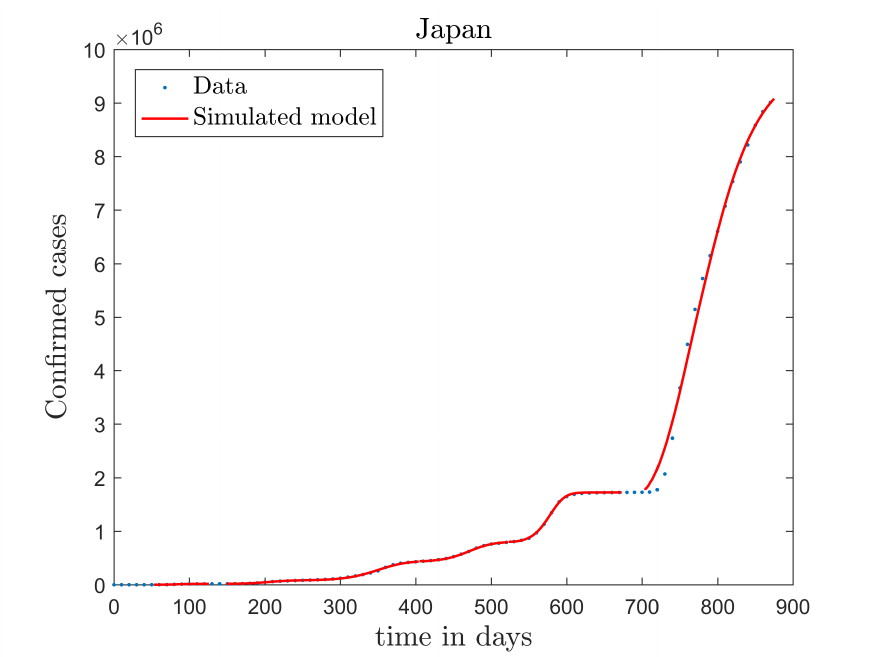}& 
		\includegraphics[scale=0.55]{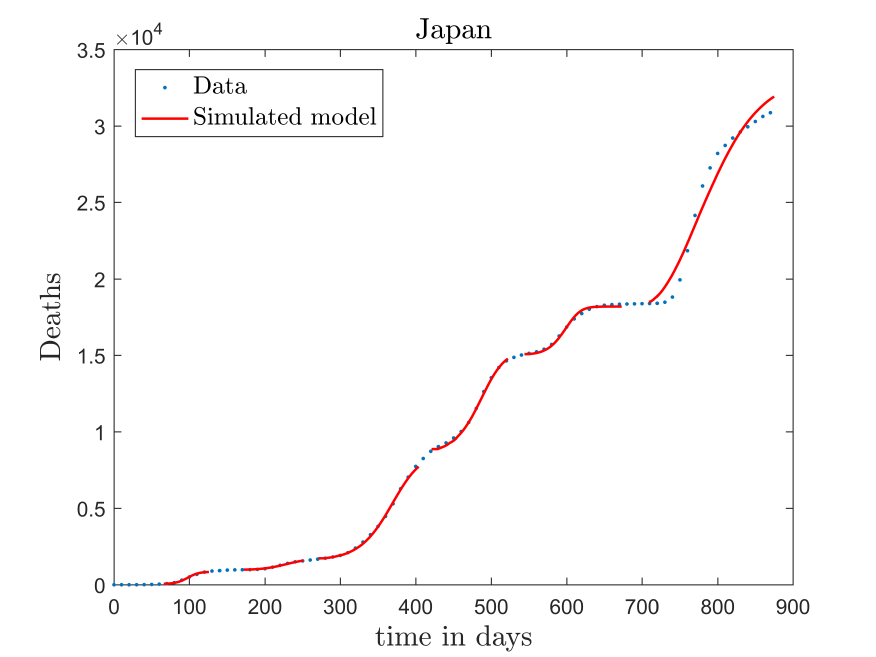}\\
		\includegraphics[scale=0.55]{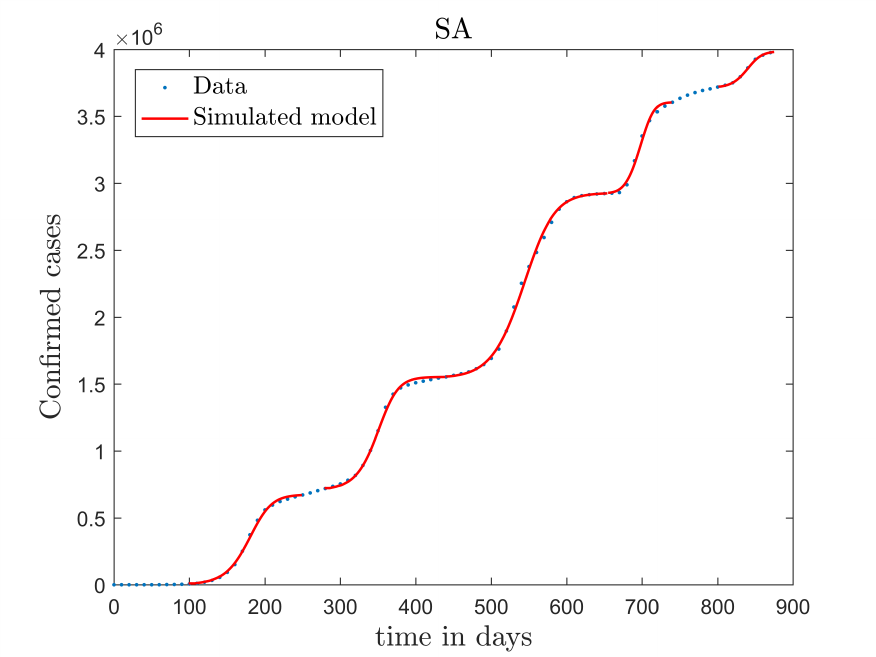}& 
		\includegraphics[scale=0.55]{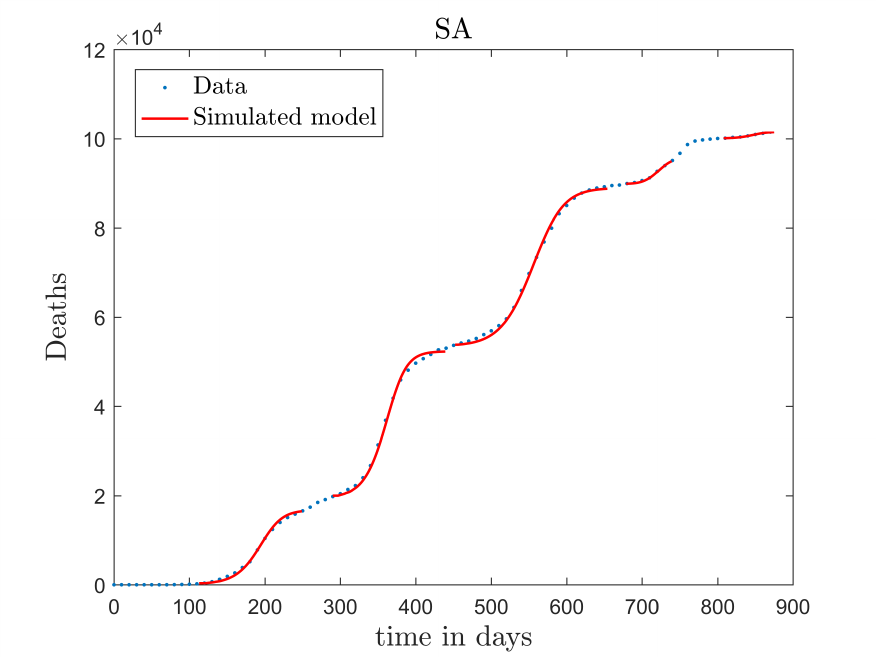}
	\end{tabular}
	
	\caption{Left column shows model fitting to the confirmed cases data of different countries. Simulated variations of deaths for different countries are shown in right column.}\label{All_country}	
\end{figure}

In Fig.~\ref{Italy_all}, we have combined all the fitted plots and re-scaled them to plot against the whole data of Italy from 22/01/2020 to 15/06/2022. Fig.~\ref{It_it} represents the plot for confirmed cases data as well as our model. Fig.~\ref{It_d} represents the same for number of deaths.

In Fig.~\ref{All_country}, we have shown variations of confirmed cases as well as deaths with time for different countries. In confirmed cases plots, there are some intermediate parts which are not fitted. These regions in-between the peaks, the daily number of new cases are very low and tending to almost zero in comparison to the earlier days. This leads to an almost flat region in the curve for the cumulative number of infected people. Our fit incorporates this behaviour as an end to the falling part of each of the peaks. We have judiciously chosen not to fit small regions only when the number of infected people reach the base level. Also, in death plots, there are some more parts which are not under the simulated regions. This is because of the delay ($\tau_{\scriptscriptstyle{\text{R}}}$) in death cases, as mentioned earlier. Thus for each peak initial $\tau_{\scriptscriptstyle{\text{R}}}-1$ number of points are omitted for simulation, as described in Eq.~\ref{death_delay}.

\begin{table}[H]
	\centering
	\begin{tabular}{|c|c|c|c|c|c|c|c|}
	\hline
	\multirow{2}{*}{Country} & \multirow{2}{*}{Parameters} & \multicolumn{6}{c|}{Peak number} \\
	\cline{3-8}
	& & 1st & 2nd & 3rd & 4th & 5th & 6th\\
	\hline
	\multirow{5}{*}{Italy} & $q$ & 0.36985 & 0.21206 & 0.24585 & 0.16389 & 0.35715 & 0.13025 \\ 
	\cline{2-8}
	& $n$ & 1.06260 & 2.82571 & 2.95011 & 2.79717 & 2.73188 & 2.21768 \\
	\cline{2-8}
	& $\tau_{\scriptscriptstyle{\text{I}}}$ & 3 & 4 & 7 & 5 & 5 & 1 \\
	\cline{2-8}
	& $\tau_{\scriptscriptstyle{\text{R}}}$ & 4 & 17 & 9 & 20 & 10 & 12\\
	\cline{2-8}
	& $I(0)$ & 111 & 151 & 362 & 227 & 162 & 254 \\
	\hhline{|========|}
	\multirow{5}{*}{Germany} & $q$ & 0.20273 & 0.10106 & 0.16616 & 0.17170 & 0.18498 & 0.11204\\ 
	\cline{2-8}
	& $n$ & 1.09335 & 2.81062 & 1.12124 & 2.79799 & 1.59799 & 2.31692 \\
	\cline{2-8}
	& $\tau_{\scriptscriptstyle{\text{I}}}$ & 3 & 7 & 9 & 4 & 2 & 5 \\
	\cline{2-8}
	& $\tau_{\scriptscriptstyle{\text{R}}}$ & 14 & 36 & 14 & 20 & 10 & 20 \\
	\cline{2-8}
	& $I(0)$ & 96 & 228 & 397 & 144 & 99 & 203\\
	\hhline{|========|}
	\multirow{5}{*}{France} & $q$ & 0.37132 & 0.16421 & 0.11439 & 0.17635 & 0.25612 & 0.14976 \\ 
	\cline{2-8}
	& $n$ & 1.57481 & 0.45996 & 0.93457 & 2.79593 & 2.65914 & 2.85849 \\
	\cline{2-8}
	& $\tau_{\scriptscriptstyle{\text{I}}}$ & 4 & 9 & 8 & 4 & 3 & 1 \\
	\cline{2-8}
	& $\tau_{\scriptscriptstyle{\text{R}}}$ & 7 & 18 & 12 & 18 & 12 & 17 \\
	\cline{2-8}
	& $I(0)$ & 114 & 176 & 170 & 196 & 218 & 481 \\
	\hhline{|========|}
	\multirow{5}{*}{US} & $q$ & 0.19558 & 0.11801 & 0.17678 & 0.17359 & -- & --\\ 
	\cline{2-8}
	& $n$ & 2.86244 & 2.60631 & 2.92009 & 2.17452 & -- & --\\
	\cline{2-8}
	& $\tau_{\scriptscriptstyle{\text{I}}}$ & 5 & 5 & 5 & 2 & -- & --\\
	\cline{2-8}
	& $\tau_{\scriptscriptstyle{\text{R}}}$ & 11 & 18 & 21 & 19 & --& --\\
	\cline{2-8}
	& $I(0)$ & 156 & 97 & 92 & 166 & --& --\\
	\hhline{|========|}
	\multirow{5}{*}{Brazil} & $q$ & 0.15814 & 0.10295 & 0.27827 & --& --& --\\ 
	\cline{2-8}
	& $n$ & 0.15893 & 2.17049 & 2.36806 & --& --& --\\
	\cline{2-8}
	& $\tau_{\scriptscriptstyle{\text{I}}}$ & 3 & 8 & 5 & --& --& --\\
	\cline{2-8}
	& $\tau_{\scriptscriptstyle{\text{R}}}$ & 8 & 16 & 11 & --& --& --\\
	\cline{2-8}
	& $I(0)$ & 11 & 62 & 135 &  --& --& --\\
	\hhline{|========|}
	\multirow{5}{*}{India} & $q$ & 0.13576 & 0.12967 & 0.30007 & --& --& --\\ 
	\cline{2-8}
	& $n$ & 1.15718 & 0.93383 & 1.84286 & --& --& --\\
	\cline{2-8}
	& $\tau_{\scriptscriptstyle{\text{I}}}$ & 8 & 3 & 2 & --& --& --\\
	\cline{2-8}
	& $\tau_{\scriptscriptstyle{\text{R}}}$ & 11 & 15 & 12 & --& --& --\\
	\cline{2-8}
	& $I(0)$ & 29 & 26 & 45 &  --& --& --\\
	\hhline{|========|}
	\multirow{5}{*}{Japan} & $q$ & 0.19671 & 0.17511 & 0.12692 & 0.14812 & 0.23149 & 0.28547 \\ 
	\cline{2-8}
	& $n$ & 1.44642 & 2.76158 & 1.58285 & 1.19748 & 2.09850 & 2.69654 \\
	\cline{2-8}
	& $\tau_{\scriptscriptstyle{\text{I}}}$ &1 & 4 & 8 & 8 & 6 & 2\\
	\cline{2-8}
	& $\tau_{\scriptscriptstyle{\text{R}}}$ & 13 & 23 & 19 & 16 & 21 & 6 \\
	\cline{2-8}
	& $I(0)$ & 123 & 132 & 73 & 200 &  96 & 42 \\
	\hhline{|========|}
	\multirow{5}{*}{SA} & $q$ & 0.17435 & 0.21300 & 0.15079 & 0.22935 & 0.29409 & --\\ 
	\cline{2-8}
	& $n$ & 1.15340 & 1.29217 & 1.60608 & 2.43431 & 1.63147 & --\\
	\cline{2-8}
	& $\tau_{\scriptscriptstyle{\text{I}}}$ & 8 & 7 & 4 & 2 & 3 & --\\
	\cline{2-8}
	& $\tau_{\scriptscriptstyle{\text{R}}}$ & 14 & 11 & 12 & 24 & 7 & --\\
	\cline{2-8}
	& $I(0)$ & 64 & 67 & 22 &  71 & 118 & --\\
	\hline
	\end{tabular}
	\caption{Best fit parameter values of different peaks of different countries.} \label{par_table}
\end{table}

\begin{figure}[H]
	\begin{subfigure}{.5\textwidth}
		\centering
		\includegraphics[scale=0.65]{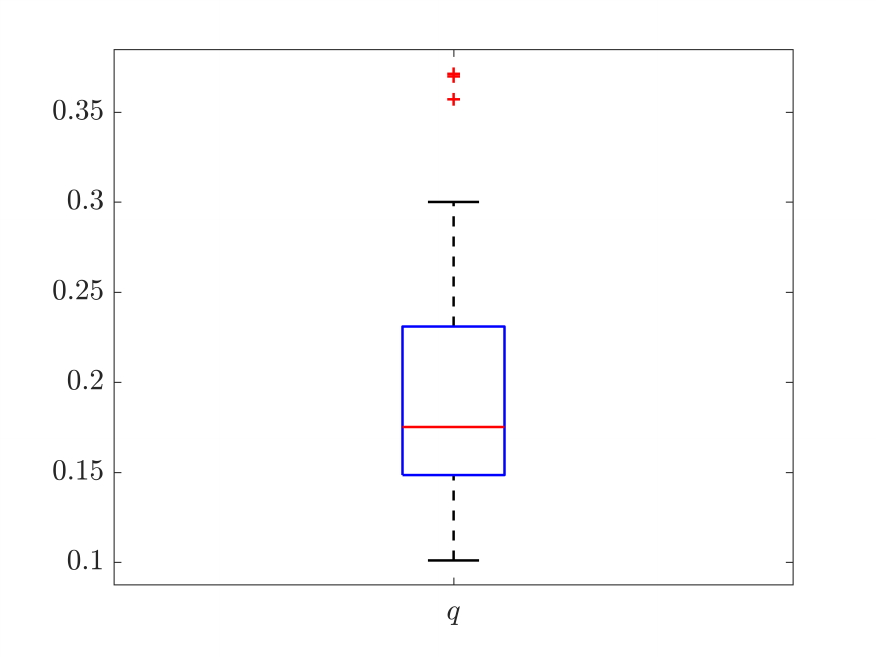}
		\caption{}
		\label{box_q}
	\end{subfigure}
	\begin{subfigure}{.5\textwidth}
		\centering
		\includegraphics[scale=0.65]{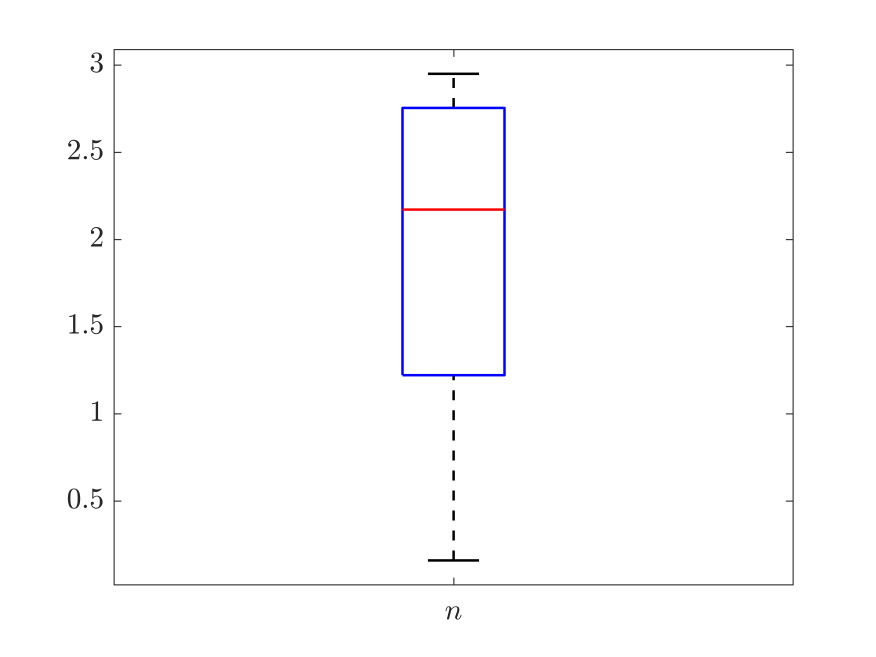}
		\caption{}
		\label{box_n}
	\end{subfigure}
	\begin{subfigure}{.5\textwidth}
		\centering
		\includegraphics[scale=0.65]{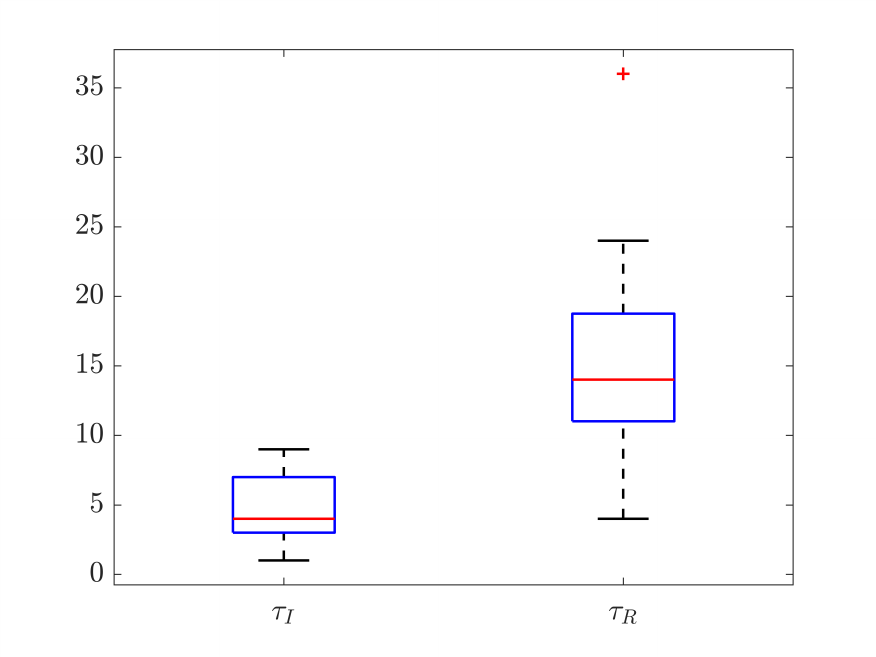}
		\caption{}
		\label{box_time}
	\end{subfigure}
	\begin{subfigure}{.5\textwidth}
		\centering
		\includegraphics[scale=0.65]{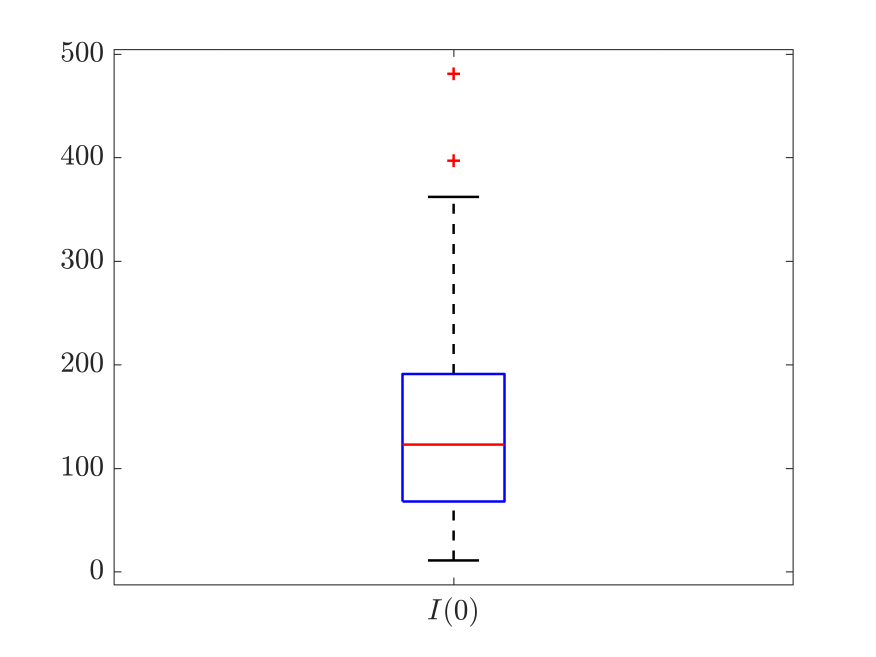}
		\caption{}
		\label{box_I}
	\end{subfigure}
	\caption{Box plots of the best-fit parameter values of different peaks of different countries. (a) Box plot of $q$, (b) box plot of $n$, (c) box plots of $\tau_{\scriptscriptstyle{\text{I}}}$ and $\tau_{\scriptscriptstyle{\text{R}}}$, (d) box plot of initial infectious cases ($I(0)$). }\label{box_plots}	
\end{figure}

Tab.~\ref{par_table} shows all the best fit parameter values to the confirmed cases data (scaled) of different peaks for various countries. The fitted values of a particular parameter are not only varied for different countries but also varied between peaks. The distribution of the best-fit values of the parameters are represented with box plots and shown in Fig.~\ref{box_plots}. In the box plot, the middle red line represents the median value of the distribution. Upper and lower edge of the box represents third and first quartile values of the distribution respectively, whereas the length of the box represents inter-quartile range (IQR). Also the upper and lower whiskers represent the values which are lain above and below the third and first quartile values respectively and have a length of 1.5*IQR. Outliers are represented with red cross. The median values of $q$, $n$, $\tau_{\scriptscriptstyle{\text{I}}}$, $\tau_{\scriptscriptstyle{\text{R}}}$ and $I(0)$ are 0.17511, 2.1705, 4, 14 and 123 respectively. From fig.~\ref{box_n}, we can see that the values of $n$ is closely distributed to 3.0.

\begin{table}[H]
	\centering
	\begin{tabular}{|c|c|c|c|c|c|c|}
		\hline
		\multirow{2}{*}{Country} & \multicolumn{6}{c|}{Peak number} \\
		\cline{2-7}
		& 1st & 2nd & 3rd & 4th & 5th & 6th\\
		\hline
		Italy  & 0.14365 & 0.02494 & 0.02080 & 0.00890 & 0.00259 & 0.00223 \\ 
		\hline
		Germany  & 0.04838 & 0.03428 & 0.01427 & 0.00614 & 0.00654 & 0.00128 \\ 
		\hline
		France  & 0.15821 & 0.01343 & 0.01644 & 0.00392 & 0.00114 & 0.00112 \\ 
		\hline
		US  & 0.01831 & 0.01463 & 0.01314 & 0.00502 & -- & --\\ 
		\hline
		Brazil  & 0.03332 & 0.02784 & 0.00550 & -- & -- & --\\ 
		\hline
		India  & 0.01540 & 0.01307 & 0.00431 & -- & -- & --\\ 
		\hline
		Japan  & 0.05050 & 0.01076 & 0.01883 & 0.01773 & 0.00338 & 0.00187 \\ 
		\hline
		SA  & 0.02473 & 0.03896 & 0.02562 & 0.00831 & 0.00538 & --\\ 
		\hline
	\end{tabular}
	\caption{Best fit values of $k$ for different peaks of different countries.} \label{k_table}
\end{table}

The values of $k$ (as described in equation.~\ref{death_rel}) which are estimated to compare our model to the deceased cases data for different peaks of various countries are shown in Tab.~\ref{k_table}. This table shows an overall decrement of $k$ values as peak number increases for a country. This means, the death rate due to COVID-19 is decreased with time and it is true for all countries.

\section{Conclusions}\label{conclusion}
To conclude, we summarize the main results and features of this work with a short discussion. Our main aim, as stated earlier, is to build a generalized probabilistic cellular automata model with the help of basic SEIR model to study the spread of epidemics, with reference to COVID-19. In this study, our main motive is to show that merging a simple model like SEIR model with a CA model with modified neighbourhood condition can produced very good results for a pandemic which is versatile in many different fronts. We have tried to show that a simple model like the one that we have adopted has been able to take into account the effects produce by the different factors of this pandemic like infectivity of the virus, social restrictions, immunity due to vaccination, re-infection and mortality rate to a large extent. Here we have described a method to fit the COVID-19 data for different peaks with different intensities and time periods. We have compared our model with some general models and tried to understand their differences. We have also fitted our model to the data for total infected cases of eight countries and tried to understand the relevance of this model in the spread of infection of COVID-19 in the different countries widely. To achieve this, we have taken the following steps:
\begin{itemize}
	\item A square lattice of the size $N\times N$ is considered for our study. We have chosen different values of $N$ like 101, 201, 301. 
	
	\item We have built a neighbourhood criteria and made some assumptions which are described in Sec.~\ref{sec2}. According to this neighbourhood criteria, the distance between any two cells ($d$) is proportional to the number of layers ($l$) between them. Also, the probability of interaction between two cells, separated by distance $d$, $p_{\scriptscriptstyle{\text{int}}}(d)\propto \frac{1}{d^{n}}$, where $n$ is power law exponent.  
	
	\item The average distance of interaction ($\langle d\rangle$) is calculated by considering the above mentioned neighbourhood condition (described in Sec.~\ref{sec3}). average distance of interaction falls very rapidly to one for positive values of $n$ and saturates to a value equal to the maximum layer number ($L$) for negative values of $n$. Also, in this section we have reasoned about the irrelevance of large positive and negative values of $n$.
	
	\item In Sec.~\ref{prb_inf} and \ref{an_study},  $Q_{\text{I}}(i,j)$, which is the probability of infection for any cell $(i,j)$, has been defined and a mathematical analysis based on this quantity $Q_{\text{I}}$ has been undertaken. We find that initially, if there are $n_{\scriptscriptstyle{\text{I}}}$ number of initial infectious cases, the newly infected cases will be $\sim n_{\scriptscriptstyle{\text{I}}}q$, where $q$ represents infectivity of the disease. Also, we have defined an equivalent quantity for the basic reproduction number ($R_{0}$) as, $q\tau_{\scriptscriptstyle{\text{R}}}$, where $\tau_{\scriptscriptstyle{\text{R}}}$ represents infectious period of the disease.  
	
	\item A suitable epidemic algorithm is defined for our model which is described in Sec.~\ref{algo}.
	
\end{itemize}

 We have run model simulations and compared our results with various other models, which are listed below:

\begin{itemize}
	\item We have simulated our model with different lattice and sample sizes (discussed in Sec.~\ref{sim_sec}). However, we could not find any significant variations in the results by changing lattice and sample sizes. Thus throughout our study, we have considered lattice size $\equiv N\times N = 101\times 101$ and sample size $S=50$.
	
	\item There are two parameters of our model which mainly control the rate of the disease spread or in other words sharpness of the infectious cases (or active cases) curve. These two parameters are: $q$, which is referred as infectivity of the disease and $n$, which is defined as `social confinement'. Thus if $q$ decrease or $n$ increase, the rate of the disease spread will decrease and the curves of the infectious cases will be flatten out. These results are portrayed in Fig.~\ref{q_and_n_var} of Sec.~\ref{sim_sec}.
	
	\item In Sec.~\ref{sim_sec}, we have shown the effect of $n$ on the spatial distribution of disease spread. As $n$ increases that is the parameter controls `social confinement', the disease spreads in small pockets in clustered form.
	
	\item We have seen that our model has been able to reproduce the basic features of the SEIR model quite well with some discrepancies as shown in Fig.~\ref{cont_v_sto}. 
	
	\item We have also compared our model with other CA models having different neighbourhood conditions. We have shown that for $n\gg 3$ our model behaves exactly similar to the models with Moore's neighbourhood condition. If we assume $n$ as a step function of layer number ($\ell$) instead of choosing $n$ is a constant, our model exhibits similar results as the CA models consisting different $r$-neighbourhood conditions. (detailed discussion in Sec.~\ref{ngh_comp_sec})
\end{itemize}

Finally, our model is used to study the time variation of COVID-19 cases in different countries for understanding this pandemic properly. We have used data of COVID-19 from eight countries, namely Italy, Germany, France, US, Brazil, India, Japan and South Africa, to obtain a fit of our model. Procedures and results of this study are described below:

\begin{itemize}
	\item We have fitted each of the peaks in the COVID-19 data separately to get the best fit parameters for our model.

	\item The data of COVID-19 for different countries is obtained from JHU CSSE data repository. Before analyzing, a proper scaling has been done on this data. Our model is fitted with the data of confirmed cases for COVID-19. To get the best fit parameters, we have minimized the sum of squared error (SSE) between the CA model and data by applying a genetic algorithm (GA), as discussed in Sec.~\ref{cov_fit_sec}.
		
	\item The parameter values obtained from the fitting is then re-checked with the data for deceased cases. Our model can generate the data for removed cases ($\equiv$ recovered+dead). However, the data for recovered cases is not fully available in JHU CSSE data repository. We have assumed that for a particular peak, deceased cases vary linearly with removed cases, where $k$ is the proportionality constant and have values between 0 and 1. This assumption help us to generate the data for deceased cases from our model and compare with the real data (detailed discussion in Sec.~\ref{cov_fit_sec}).
	
	\item The results of the fitting to the data for confirmed cases and comparison to the data for deceased cases of Italy are shown in Fig.~\ref{Italy_sep_itot}, \ref{Italy_sep_d}, \ref{Italy_all} and \ref{All_country}. Also, the optimized parameter values for different peaks of the different countries are listed in Tab.~\ref{par_table}.
	
	\item These plots and the table show that the rate of the spread of COVID-19 mainly depends on two parameters $q$ and $n$. 
	
	\item The data for deceased cases and simulated results show deviations in some situations, as seen in Fig.~\ref{All_country}. The main reason behind this is the assumption of the linear relationship between the deceased cases and the removed cases for a particular peak. This zeroth order model with a linear relationship can be seen to be valid for most situations. 
	
	\item We have drawn box plots (Fig.~\ref{box_plots}) for fitting parameters to show their distribution. An important thing to note that the median value of $n$ is at 2.1705, which is above the central value, $n=1.5$ of our chosen range. Thus theoretically we can say that the degree of `social confinement' in different countries are above average. Effect of various restrictions like social-distancing, and lockdown may be causes of this.
	
	\item Also Tab.~\ref{k_table} shows that the value of $k$ decreases (not always monotonically) as COVID-19 pandemic proceeds. This means the death rate decreases with time 
\end{itemize}

In this work, we have analyzed COVID-19 data from a different perspective. In future, we aim to compare and validate this stochastic model with physically motivated models. We hope our work will help to understand and forecast future pandemic situations. 

\section{Acknowledgment}
 One of the authors (S. C.) would like to acknowledge the financial support lent by the University Grant Commission (UGC), Government of India, in the form of CSIR-UGC NET-JRF stipend. Another author (S. R. C.) would like to thank St. Xavier's College (Autonomous), Kolkata for providing financial support in the form of `Intramural Research Grant for R\&D Projects'. S. C. would also like to thank Mr. Saswata Das and Mr. Bappaditya Manna for providing support during this work. Finally, all the authors would like to acknowledge St. Xavier’s College (Autonomous), Kolkata as their host institution.

\bibliographystyle{unsrt}
\bibliography{References}	

\begin{thebibliography}{10}

\bibitem{future_1}
Marco Marani, Gabriel~G. Katul, William~K. Pan, and Anthony~J. Parolari.
\newblock Intensity and frequency of extreme novel epidemics.
\newblock {\em Proceedings of the National Academy of Sciences},
  118(35):e2105482118, 2021.
\newblock DOI: 10.1073/pnas.2105482118.

\bibitem{future_2}
Camilo Mora, Tristan McKenzie, Isabella~M Gaw, Jacqueline~M Dean, Hannah von
  Hammerstein, Tabatha~A Knudson, Renee~O Setter, Charlotte~Z Smith, Kira~M
  Webster, Jonathan~A Patz, et~al.
\newblock Over half of known human pathogenic diseases can be aggravated by
  climate change.
\newblock {\em Nature climate change}, 12(9):869--875, 2022.
\newblock DOI: 10.1038/s41558-022-01426-1.

\bibitem{worldometers}
{\em Live update of COVID-19 situation in different countries- Worldometers}.
\newblock \url{https://www.worldometers.info/coronavirus/}.

\bibitem{WHO}
{\em Tracking SARS-CoV-2 variants (WHO)}.
\newblock
  \url{https://www.who.int/en/activities/tracking-SARS-CoV-2-variants/}.

\bibitem{NP_1}
Hannah Ritchie, Nectar Gan, Simone McCarthy, Selina Wang, and Mengchen Zhang.
\newblock Leaked notes from chinese health officials estimate 250 million
  covid-19 infections in december: reports.
\newblock {\em CNN International}, Dec, 23, 2022.

\bibitem{NP_2}
James Glanz, Mara Hvistendahl, and Agnes Chang.
\newblock How deadly was china’s covid wave?
\newblock {\em The New York Times}, Feb, 15, 2023.

\bibitem{basic_SIR}
William~Ogilvy Kermack, A.~G. McKendrick, and Gilbert~Thomas Walker.
\newblock A contribution to the mathematical theory of epidemics.
\newblock {\em Proceedings of the Royal Society of London. Series A, Containing
  Papers of a Mathematical and Physical Character}, 115(772):700, 1927.
\newblock DOI: 10.1098/rspa.1927.0118.

\bibitem{SIR_dengue}
Syafruddin Side and Mohd Salmi~Md Noorani.
\newblock A sir model for spread of dengue fever disease (simulation for south
  sulawesi, indonesia and selangor, malaysia).
\newblock {\em World Journal of Modelling and Simulation}, 9(2):96--105, 2013.

\bibitem{SIRS_influenza}
Mevin~B. Hooten, Jessica Anderson, and Lance~A. Waller.
\newblock Assessing north american influenza dynamics with a statistical sirs
  model.
\newblock {\em Spatial and Spatio-temporal Epidemiology}, 1(2):177--185, 2010.
\newblock DOI: 10.1016/j.sste.2010.03.003.

\bibitem{Rep_No_Influenza}
Daisuke Furushima, Shoko Kawano, Yuko Ohno, and Masayuki Kakehashi.
\newblock Estimation of the basic reproduction number of novel influenza a
  (h1n1) pdm09 in elementary schools using the sir model.
\newblock {\em The Open Nursing Journal}, 11:64, 2017.
\newblock DOI: 10.2174/1874434601711010064.

\bibitem{SEIAR_influenza}
Tianmu Chen, Yuanxiu Huang, Ruchun Liu, Zhi Xie, Shuilian Chen, and Guoqing Hu.
\newblock Evaluating the effects of common control measures for influenza a
  (h1n1) outbreak at school in china: A modeling study.
\newblock {\em PLOS ONE}, 12(5):1--20, 2017.
\newblock DOI: 10.1371/journal.pone.0177672.

\bibitem{SEIR_ebola}
Christian~L Althaus.
\newblock Estimating the reproduction number of ebola virus (ebov) during the
  2014 outbreak in west africa.
\newblock {\em PLoS currents},
  6:ecurrents.outbreaks.91afb5e0f279e7f29e7056095255b288, 2014.
\newblock DOI: 10.1371/currents.outbreaks.91afb5e0f279e7f29e7056095255b288.

\bibitem{Ebola_afr}
T.~Berge, J.M.-S. Lubuma, G.M. Moremedi, N.~Morris, and R.~Kondera-Shava.
\newblock A simple mathematical model for ebola in africa.
\newblock {\em Journal of Biological Dynamics}, 11(1):42--74, 2017.
\newblock DOI: 10.1080/17513758.2016.1229817.

\bibitem{cont_bio_model_measles}
K.~Glass, Y.~Xia, and B.T. Grenfell.
\newblock Interpreting time-series analyses for continuous-time biological
  models measles as a case study.
\newblock {\em Journal of Theoretical Biology}, 223(1):19 -- 25, 2003.
\newblock DOI: 10.1016/s0022-5193(03)00031-6.

\bibitem{covid_vac_india}
Brody~H. Foy, Brian Wahl, Kayur Mehta, Anita Shet, Gautam~I. Menon, and Carl
  Britto.
\newblock Comparing covid-19 vaccine allocation strategies in india: A
  mathematical modelling study.
\newblock {\em International Journal of Infectious Diseases}, 103:431, 2021.
\newblock DOI: 10.1016/j.ijid.2020.12.075.

\bibitem{covid_vac_acc_india}
Namrata Soni, Jyoti Bhola, Ashutosh Yadav, Ishita Srivastva, and Utcarsh
  Mathur.
\newblock A mathematical reflection of covid-19 and vaccination acceptance in
  india.
\newblock 8:150, 2021.
\newblock DOI: 10.21276/apjhs.2021.8.3.27.

\bibitem{NPI_vs_vac}
Ting-Yu Lin, Sih-Han Liao, Chao-Chih Lai, Eugenio Paci, and Shao-Yuan Chuang.
\newblock Effectiveness of non-pharmaceutical interventions and vaccine for
  containing the spread of covid-19: Three illustrations before and after
  vaccination periods.
\newblock {\em Journal of the Formosan Medical Association}, 120:S46, 2021.
\newblock DOI: 10.1016/j.jfma.2021.05.015.

\bibitem{vac_resistant}
Simon~A Rella, Yuliya~A Kulikova, Emmanouil~T Dermitzakis, and Fyodor~A
  Kondrashov.
\newblock Rates of sars-cov-2 transmission and vaccination impact the fate of
  vaccine-resistant strains.
\newblock {\em Scientific Reports}, 11(1):15729, 2021.
\newblock DOI: 10.1038/s41598-021-95025-3.

\bibitem{HIT}
Sourav Chowdhury, Suparna Roychowdhury, and Indranath Chaudhuri.
\newblock Universality and herd immunity threshold: Revisiting the sir model
  for covid-19.
\newblock {\em International Journal of Modern Physics C}, 32(10):2150128,
  2021.
\newblock DOI: 10.1142/S012918312150128X.

\bibitem{3rdwave_covid}
Sourav Chowdhury, Suparna Roychowdhury, and Indranath Chaudhuri.
\newblock A robust prediction from a minimal model of covid-19--can we avoid
  the third wave?
\newblock {\em arXiv preprint arXiv:2112.08794}, 2021.

\bibitem{chaos_vac}
Andr\'e~F. Steklain, Ahmed Al-Ghamdi, and Euaggelos~E. Zotos.
\newblock Using chaos indicators to determine vaccine influence on epidemic
  stabilization.
\newblock {\em Phys. Rev. E}, 103:032212, Mar 2021.
\newblock DOI: 10.1103/PhysRevE.103.032212.

\bibitem{chaos_vac_osc}
Jorge Duarte, Cristina Janu{\'a}rio, Nuno Martins, Jes{\'u}s Seoane, and
  Miguel~AF Sanju{\'a}n.
\newblock Controlling infectious diseases: the decisive phase effect on a
  seasonal vaccination strategy.
\newblock {\em arXiv preprint arXiv:2102.08284}, 2021.

\bibitem{chikungunya_vector_reg2}
Gerardo Ortigoza, Fred Brauer, and Iris Neri.
\newblock Modelling and simulating chikungunya spread with an unstructured
  triangular cellular automata.
\newblock {\em Infectious Disease Modelling}, 5:197--220, 2020.
\newblock DOI: 10.1016/j.idm.2019.12.005.

\bibitem{dengue_CA}
Puspa Eosina, Taufik Djatna, and Helda Khusun.
\newblock A cellular automata modeling for visualizing and predicting spreading
  patterns of dengue fever.
\newblock {\em TELKOMNIKA}, 14(1):228, 2016.
\newblock DOI: 10.12928/TELKOMNIKA.v14i1.2404.

\bibitem{Moore}
S.~Hoya White, A.~Martín {del Rey}, and G.~Rodríguez Sánchez.
\newblock Modeling epidemics using cellular automata.
\newblock {\em Applied Mathematics and Computation}, 186(1):193--202, 2007.
\newblock DOI: 10.1016/j.amc.2006.06.126.

\bibitem{Moore_2}
Gui-Quan Sun, Zhen Jin, Li-Peng Song, Amit Chakraborty, and Bai-Lian Li.
\newblock Phase transition in spatial epidemics using cellular automata with
  noise.
\newblock {\em Ecological research}, 26:333--340, 2011.
\newblock 10.1007/s11284-010-0789-9.

\bibitem{Moore_3}
B.~Cissé, S.~El Yacoubi, and A.~Tridane.
\newblock Impact of neighborhood structure on epidemic spreading by means of
  cellular automata approach.
\newblock {\em Procedia Computer Science}, 18:2603--2606, 2013.
\newblock 2013 International Conference on Computational Science.

\bibitem{influenza_seir_reg1}
A.~Holko, M.~Medrek, Z.~Pastuszak, and K.~Phusavat.
\newblock Epidemiological modeling with a population density map-based cellular
  automata simulation system.
\newblock {\em Expert Systems with Applications}, 48:1--8, 2016.
\newblock DOI: 10.1016/j.eswa.2015.08.018.

\bibitem{influenza_egypt}
Khaled~M Khalil, M~Abdel-Aziz, Taymour~T Nazmy, and Abdel-Badeeh~M Salem.
\newblock An agent-based modeling for pandemic influenza in egypt.
\newblock In {\em Handbook on Decision Making}, volume~33, pages 205--218.
  Springer, 2012.
\newblock DOI: 10.1007/978-3-642-25755-1\_11.

\bibitem{CA_influenza_abu_dhabi}
Senthil Athithan, Vidya~Prasad Shukla, and Sangappa~Ramachandra Biradar.
\newblock Epidemic spread modeling with time variant infective population using
  pushdown cellular automata.
\newblock {\em Journal of Computational Environmental Sciences}, 2014:769064,
  2014.
\newblock DOI: 10.1155/2014/769064.

\bibitem{influenzaA_SLEIRD}
Sheng Bin, Gengxin Sun, and Chih-Cheng Chen.
\newblock Spread of infectious disease modeling and analysis of different
  factors on spread of infectious disease based on cellular automata.
\newblock {\em International Journal of Environmental Research and Public
  Health}, 16(23), 2019.
\newblock DOI: 10.3390/ijerph16234683.

\bibitem{dengue_vector}
Murali~Krishna Enduri and Shivakumar Jolad.
\newblock Dynamics of dengue disease with human and vector mobility.
\newblock {\em Spatial and Spatio-temporal Epidemiology}, 25:57--66, 2018.
\newblock DOI: 10.1016/j.sste.2018.03.001.

\bibitem{Ebola_CA}
Emily Burkhead and Jane Hawkins.
\newblock A cellular automata model of ebola virus dynamics.
\newblock {\em Physica A: Statistical Mechanics and its Applications},
  438:424--435, 2015.
\newblock DOI: 10.1016/j.physa.2015.06.049.

\bibitem{CA_nghbd_sat}
Armin~R. Mikler, Sangeeta Venkatachalam, and Kaja Abbas.
\newblock Modeling infectious diseases using global stochastic cellular
  automata.
\newblock {\em Journal of Biological Systems}, 13(04):421--439, 2005.
\newblock DOI: 10.1142/S0218339005001604.

\bibitem{CA_small_world}
Henrique~Fabricio Gagliardi and Domingos Alves.
\newblock Small-world effect in epidemics using cellular automata.
\newblock {\em Mathematical Population Studies}, 17(2):79--90, 2010.
\newblock DOI: 10.1080/08898481003689486.

\bibitem{covid_social_isol}
P.H.T. Schimit.
\newblock A model based on cellular automata to estimate the social isolation
  impact on covid-19 spreading in brazil.
\newblock {\em Computer Methods and Programs in Biomedicine}, 200:105832, 2021.
\newblock DOI: 10.1016/j.cmpb.2020.105832.

\bibitem{covid_vac_lockdwn}
P.K. Jithesh.
\newblock A model based on cellular automata for investigating the impact of
  lockdown, migration and vaccination on covid-19 dynamics.
\newblock {\em Computer Methods and Programs in Biomedicine}, 211:106402, 2021.
\newblock DOI: 10.1016/j.cmpb.2021.106402.

\bibitem{covid_GA_1}
Sayantari Ghosh and Saumik Bhattacharya.
\newblock A data-driven understanding of covid-19 dynamics using sequential
  genetic algorithm based probabilistic cellular automata.
\newblock {\em Applied Soft Computing}, 96:106692, 2020.
\newblock DOI: 10.1016/j.asoc.2020.106692.

\bibitem{covid_mob_restr_net}
L.~L. Lima and A.~P.~F. Atman.
\newblock Impact of mobility restriction in covid-19 superspreading events
  using agent-based model.
\newblock {\em PLOS ONE}, 16(3):1--17, 03 2021.
\newblock DOI: 10.1371/journal.pone.0248708.

\bibitem{COVID_GA_SC}
Sourav Chowdhury, Suparna Roychowdhury, and Indranath Chaudhuri.
\newblock Cellular automata in the light of covid-19.
\newblock {\em The European Physical Journal Special Topics},
  231(18-20):3619--3628, 2022.
\newblock DOI: 10.1140/epjs/s11734-022-00619-1.

\bibitem{CA_new_1}
Yusra~Bibi Ruhomally, Maheshsingh Mungur, Abdel Anwar~Hossen Khoodaruth,
  Vishwamitra Oree, and Muhammad~Zaid Dauhoo.
\newblock Assessing the impact of contact tracing, quarantine and red zone on
  the dynamical evolution of the covid-19 pandemic using the cellular automata
  approach and the resulting mean field system: A case study in mauritius.
\newblock {\em Applied Mathematical Modelling}, 111:567--589, 2022.
\newblock DOI: 10.1016/j.apm.2022.07.008.

\bibitem{CA_new_2}
Ihor Kosovych, Igor Cherevko, Denys Nevinskyi, and Yaroslav Vyklyuk.
\newblock Simulation of various distribution restrictions of covid-19 using
  cellular automata.
\newblock In {\em 2022 12th International Conference on Advanced Computer
  Information Technologies (ACIT)}, pages 58--61, 2022.
\newblock DOI: 10.1109/ACIT54803.2022.9913172.

\bibitem{CA_new_3}
Puspa Eosina, Aniati~Murni Arymurthy, and Adila~Alfa Krisnadhi.
\newblock A non-uniform continuous cellular automata for analyzing and
  predicting the spreading patterns of covid-19.
\newblock {\em Big Data and Cognitive Computing}, 6(2), 2022.
\newblock DOI: 10.3390/bdcc6020046.

\bibitem{covid_GA_2}
Larissa~M. Fraga, Gina M.~B. de~Oliveira, and Luiz G.~A. Martins.
\newblock Multistage evolutionary strategies for adjusting a cellular
  automata-based epidemiological model.
\newblock In {\em 2021 IEEE Congress on Evolutionary Computation (CEC)}, pages
  466--473, 2021.
\newblock DOI: 10.1109/CEC45853.2021.9504738.

\bibitem{sp_1}
George Grekousis.
\newblock {\em Spatial analysis methods and practice:
  describe--explore--explain through GIS}.
\newblock Cambridge University Press, 2020.

\bibitem{sp_4}
A.~Stewart Fotheringham.
\newblock Spatial structure and distance-decay parameters.
\newblock {\em Annals of the Association of American Geographers},
  71(3):425--436, 1981.

\bibitem{JHU_data}
Ensheng Dong, Hongru Du, and Lauren Gardner.
\newblock An interactive web-based dashboard to track covid-19 in real time.
\newblock {\em The Lancet infectious diseases}, 20(5):533--534, 2020.
\newblock DOI: 10.1016/S1473-3099(20)30120-1.

\end{thebibliography}

\end{document}